\documentclass[journal]{IEEEtran}

\usepackage{scalerel}
\usepackage{times}
\usepackage{relsize}

\usepackage{enumitem}
\usepackage{float}
\usepackage{mathtools}
\usepackage{graphicx}
\usepackage{graphics}
\usepackage[dvips]{epsfig}
\usepackage{subfig}
\usepackage{verbatim}
\usepackage{graphicx,float,latexsym,amssymb,amsfonts,amsmath,amstext,times,epsfig,setspace}
\usepackage{algorithm}
\usepackage[utf8]{inputenc}
\usepackage[noend]{algpseudocode}
\usepackage{fancyhdr}
\usepackage{stfloats}
\usepackage{tabu}
\usepackage{booktabs}
\usepackage{amsmath}
\usepackage[noend]{algpseudocode}
\renewcommand{\algorithmiccomment}[1]{\bgroup\hfill\scriptsize//~#1\egroup}

\DeclareMathOperator*{\argmax}{arg\,max}
\DeclareMathOperator*{\argmin}{arg\,min}

\newcommand{\Exp}[1]{\mathbf{E}\left[{#1}\right]}

\newcommand{\Prob}[1]{\mathbf{Pr}\left[{#1}\right]}
\newcommand{\Probi}[2]{\mathbf{Pr}_{#1}\left[{#2}\right]}

\newcommand{\entropy}[1]{\mathbf{H}\left({#1}\right)}

\newcommand{\liqsystem}{liquid system}
\newcommand{\Liqsystem}{Liquid system}
\newcommand{\liqrepairer}{liquid repairer}

\newcommand{\advliqrepairer}{advanced liquid repairer}

\newcommand{\tradsystem}{small code system}
\newcommand{\Tradsystem}{Small code system}

\newcommand{\tradrepair}{reactive repair}

\newcommand{\Regframe}{Regenerating framework}
\newcommand{\Reglower}{Regenerating lower bound}
\newcommand{\Regrepairer}{Regenerating repairer}
\newcommand{\Erepairer}{Equal-read repairer}

\newcommand{\srcdata}{source data}
\newcommand{\Srcdata}{Source data}

\newcommand{\Nnum}{N}
\newcommand{\erate}{{\rm \it erate}}
\newcommand{\rrate}{{\rm \it rrate}}
\newcommand{\nsize}{{\rm \it nsize}}
\newcommand{\osize}{{\rm \it osize}}
\newcommand{\rsize}{{\rm \it rsize}}
\newcommand{\ssize}{{\rm \it ssize}}
\newcommand{\rfsize}{{\rm \it rfsize}}
\newcommand{\vsz}{{\rm \it vsize}}
\newcommand{\dsize}{{\rm \it dsize}}

\newcommand{\nnum}{n}
\newcommand{\knum}{k}
\newcommand{\rnum}{r}

\newcommand{\betap}{{\beta'}}

\newcommand{\Rfnc}{{\mathbf{R}}}
\newcommand{\Rpfnc}{{\mathbf{R}'}}
\newcommand{\Afnc}{{\mathbf{A}}}

\newcommand{\fseq}{{\rm \it fseq}}
\newcommand{\Bseq}{{\rm \it Bseq}}
\newcommand{\Bseqseq}{{\rm \it Bseqseq}}

\newcommand{\Gseq}{{\rm \it Gseq}}
\newcommand{\Ghseq}{\hat{\rm \it G}{\rm \it seq}}
\newcommand{\Qseq}{{\rm \it Qseq}}
\newcommand{\tseq}{{\rm \it tseq}}
\newcommand{\Tseq}{{\rm \it Tseq}}

\newcommand{\Tpseq}{\hat{\rm T}{\rm seq}}
\newcommand{\idseq}{{\rm \it idseq}}
\newcommand{\IDseq}{{\rm \it IDseq}}
\newcommand{\Useq}{{\rm \it Useq}}

\newcommand{\Bvar}{B}
\newcommand{\Cdata}{C}
\newcommand{\dvar}{d}
\newcommand{\dhvar}{\hat{d}}
\newcommand{\Ddata}{D}
\newcommand{\Fdata}{F}

\newcommand{\Gvar}{G}
\newcommand{\Ghvar}{\hat{G}}
\newcommand{\Gsum}{{\rm \it GS}}
\newcommand{\idvar}{{\rm \it id}}
\newcommand{\IDvar}{{\rm \it ID}}
\newcommand{\Mdata}{M}
\newcommand{\Mpdata}{M'}
\newcommand{\Ppred}{P}
\newcommand{\pvar}{p}
\newcommand{\phvar}{\hat{p}}
\newcommand{\Qvar}{Q}
\newcommand{\Rdata}{R}
\newcommand{\RFdata}{{\rm \it RF}}
\newcommand{\Sdata}{S}
\newcommand{\thvar}{\hat{t}}

\newcommand{\tvar}{t}
\newcommand{\Tvar}{T}
\newcommand{\tvarp}{t^{\scaleto{+}{4pt}}}
\newcommand{\tvarm}{t^{\scaleto{-}{4pt}}}

\newcommand{\Thvar}{\hat{T}}
\newcommand{\Thvarm}{\hat{T}^{\scaleto{-}{4pt}}}
\newcommand{\Uvar}{U}
\newcommand{\Vdata}{V}
\newcommand{\xvar}{x}
\newcommand{\xpvar}{x'}

\newcommand{\Xvar}{X}
\newcommand{\Xset}{{\cal X}}
\newcommand{\Xbset}{{\bar{\cal X}}}
\newcommand{\Yvar}{Y}
\newcommand{\Ypvar}{Y'}
\newcommand{\Yppvar}{Y''}
\newcommand{\Zvar}{Z}
\newcommand{\zvar}{z}

\newcommand{\epscore}{\epsilon_{c}}
\newcommand{\delcore}{\delta_{\scriptscriptstyle \rm c}}
\newcommand{\epsgeo}{\epsilon_{d}}
\newcommand{\delgeo}{\delta_{d}}

\newcommand{\deluni}{\delta_{u}}
\newcommand{\epspoi}{\epsilon}
\newcommand{\delpoi}{\delta}

\newcommand{\rcapacity}{capacity}

\newcommand{\erasurerate}{erasure rate}

\newcommand{\storeoverhead}{storage overhead}

\newcommand{\rrepairrate}{read rate}

\newcommand{\placegroup}{placement group}
\newcommand{\Placegroup}{Placement group}

\newcommand{\len}[1]{\lvert\lvert{#1}\rvert\rvert}
\newcommand{\abs}[1]{\left\lvert{#1}\right\rvert}
\newcommand{\ang}[1]{\langle{#1}\rangle}
\newcommand{\set}[1]{\left\{{#1}\right\}}

\newcommand{\bitval}{bit}

\newcommand{\lrepairer}{local-computation repairer}
\newcommand{\Lrepairer}{Local-computation repairer}

\newcommand{\inu}{\in_{\cal U}}

\newcommand{\nfail}{failure}
\newcommand{\Nfail}{Failure}
\newcommand{\nfseq}{failure sequence}

\newcommand{\timeseq}{timing sequence}

\newcommand{\identifierseq}{identifier sequence}

\newcommand{\didentifierseq}{distinct identifier sequence}

\newcommand{\uidentifierseq}{uniform identifier sequence distribution}
\newcommand{\Uidentifierseq}{Uniform identifier sequence distribution}

\newcommand{\identifier}{identifier}

\newcommand{\Ptdist}{Poisson failure distribution}

\newcommand{\lnifunction}{{\mathbf{lni}}}
\newcommand{\lndfunction}{{\mathbf{lnd}}}

\newcommand{\expp}[1]{10^{#1}}
\newcommand{\expm}[1]{10^{- #1}}

\newtheorem{theorem}{Theorem}[section]

\newtheorem{complemma}[theorem]{{\bf Compression Lemma}}
\newcommand{\Complemma}{Compression Lemma~\ref{compression lemma}}

\newtheorem{dimlemma}[theorem]{{\bf Dimakis Lemma}}
\newcommand{\Dimlemma}{Dimakis Lemma~\ref{dim lemma}}

\newtheorem{compcorollary}[theorem]{{\bf Compression Corollary}}
\newcommand{\Compcorollary}{Compression Corollary~\ref{compression corollary}}

\newtheorem{elemma}[theorem]{{\bf Equal-read Lemma}}
\newcommand{\Elemma}{Equal-read Lemma~\ref{equal lemma}}

\newtheorem{corelemma}[theorem]{{\bf Core Lemma}}
\newcommand{\Corelemma}{Core Lemma~\ref{core lemma}}

\newtheorem{coretheorem}[theorem]{{\bf Core Theorem}}
\newcommand{\Coretheorem}{Core Theorem~\ref{core theorem}}

\newtheorem{supertheorem}[theorem]{{\bf Supermartingale Theorem}}
\newcommand{\Supertheorem}{Supermartingale Theorem~\ref{super theorem}}

\newtheorem{geolemma}[theorem]{{\bf Distinct Failures Lemma}}
\newcommand{\Geolemma}{Distinct Failures Lemma~\ref{geo lemma}}

\newtheorem{unitheorem}[theorem]
{{\bf Uniform Failures Theorem}}
\newcommand{\Unitheorem}
{Uniform Failures Theorem~\ref{uniform theorem}}

\newtheorem{Poissontheorem}[theorem]
{{\bf Poisson Failures Theorem}}
\newcommand{\Poissonthm}
{Poisson Failures Theorem~\ref{Poisson theorem}}

\newenvironment{proof}[1][Proof]{\begin{trivlist}
\item[\hskip \labelsep {\bfseries #1}]}{\end{trivlist}}

\newcommand{\qed}{\nobreak \ifvmode \relax \else
      \ifdim\lastskip<1.5em \hskip-\lastskip
      \hskip1.5em plus0em minus0.5em \fi \nobreak
      \vrule height0.75em width0.5em depth0.25em\fi}

\begin{document}

\title{Repair rate lower bounds for distributed storage}
\thispagestyle{fancy}
\author{Michael~Luby,~\IEEEmembership{~IEEE~Fellow,~ACM~Fellow} 
\IEEEcompsocitemizethanks{\IEEEcompsocthanksitem Portions of this work were done while the author was with Qualcomm Technologies, Inc.  The author is currently with International Computer Science Institute, Berkeley CA 94704 e-mail: luby@icsi.berkeley.edu, theluby@ieee.org.}
\thanks{Revised draft: July 23, 2019}
}

\fancyhead[LO]{\small Luby, Repair rate lower bounds for distributed storage}
\fancyfoot[L]{\em Revised draft: July 23, 2019}
\renewcommand{\headrulewidth}{0pt}

\maketitle

\begin{abstract}

One of the primary objectives of a distributed storage system is to reliably store 
a large amount $\dsize$ of \srcdata\ for long durations using a 
large number $\Nnum$ of unreliable storage nodes, 
each with \rcapacity\ $\nsize$.
The \storeoverhead\ $\beta$ is the fraction of system \rcapacity\ 
available beyond $\dsize$, i.e.,
\begin{equation}
\label{beta eq}
\beta = 1- \frac{\dsize}{\Nnum \cdot \nsize},
\end{equation}  
Storage nodes fail randomly over time and are replaced with initially empty nodes, 
and thus data is erased from the system at an average rate 
\begin{equation}
\label{erate eq}
\erate = \lambda \cdot \Nnum \cdot \nsize,
\end{equation}
where $1/\lambda$ is the average lifetime of a node before failure.
 
To maintain recoverability of the \srcdata, a repairer continually 
reads data over a network from nodes at some average rate $\rrate$, 
and generates and writes data to nodes based on the read data.

The main result is that, for any repairer,
if the \srcdata\ is recoverable at each point in time then  
it must be the case that
\begin{equation}
\label{rrate lb eq}
\rrate \ge \frac{\erate}{2 \cdot \beta}
\end{equation}
asymptotically as $\Nnum$ goes to infinity and $\beta$ goes to zero.
Thus, Inequality~\eqref{rrate lb eq} provides a fundamental 
lower bound on the average rate that any repairer needs to read data from the
system in order to maintain recoverability of the \srcdata.
\end{abstract}


\IEEEpeerreviewmaketitle

\section{Overview}
\label{overview sec}

\IEEEPARstart{A}{ }distributed storage system generically consists of 
interconnected storage nodes, where each node can store a large quantity of data.
We let $\Nnum$ be the number of storage nodes in the system, where
each node has $\nsize$ \bitval s of storage capacity. 

Commonly, distributed storage systems are built using relatively inexpensive
and generally not completely reliable hardware.
For example, nodes can go offline for periods of time (transient failure), 
in which case the data they store is temporarily unavailable, 
or permanently fail, in which case the data they store is permanently erased.   
Permanent \nfail s are not uncommon, and transient \nfail s are frequent.

Although it is often hard to accurately model \nfail s, 
an independent \nfail\ model can provide insight into 
the strengths and weaknesses of a practical system, 
and can provide a first order approximation to how a practical system operates.
In fact, one of the primary reasons practical storage systems 
are built using distributed infrastructure is so that failures of 
the infrastructure are as independent as possible.

In our model, each storage node permanently fails independently
and randomly at rate $\lambda$
at each point in time and is replaced with a new node initialized 
to zeroes when it fails, and thus \bitval s are erased from the 
system at an average rate $\erate$ as defined in Equation~\eqref{erate eq}.

A primary goal of a distributed storage system is to reliably store as much
\srcdata\ as possible for a long time, 
i.e., at each point in time the \srcdata\ should be recoverable from the data
stored in the system at that point in time.  We let $\dsize$ be the size of the
\srcdata\ to be stored.  To maintain recoverability of the \srcdata,
a {\em repairer} continually reads data over a network from nodes at 
some average rate $\rrate$, and generates and writes data 
to nodes based on the read data.

Distributed storage systems generally allocate a 
fraction of their \rcapacity\ to \storeoverhead,
which is used by the repairer to help maintain 
recoverability of \srcdata\ as \nfail s occur.
The {\em \storeoverhead} $\beta$ is the fraction of 
\rcapacity\ available beyond the size of the \srcdata, i.e.,
$\beta$ is defined in Equation~\eqref{beta eq},
and thus $\dsize = (1-\beta) \cdot \Nnum \cdot \nsize$.

The main result is that, for any repairer,
if the \srcdata\ is recoverable at each point in time then  
it must be the case that Inequality~\eqref{rrate lb eq} holds
asymptotically as $\Nnum$ goes to infinity and $\beta$ goes to zero.
Thus, Inequality~\eqref{rrate lb eq} provides a fundamental 
lower bound on the average rate that any repairer needs to read data from the
system in order to maintain recoverability of the \srcdata.
  
The repairers described in~\cite{LubyRich19} have a peak \rrepairrate\
that is at most the righthand size of Inequality~\eqref{rrate lb eq} asymptotically 
as $\Nnum$ goes to infinity and $\beta$ goes to zero, 
and thus
\[ \rrate = \frac{\erate}{2 \cdot \beta} \]
expresses a fundamental trade-off between the repairer \rrepairrate\ 
and \storeoverhead\ as a function of the \erasurerate. 

\subsection{Practical system parameters}
\label{practical setting sec}

An example of a practical system is one with $\Nnum = \expp{5}$ nodes, 
with $\nsize = \expp{16}$ \bitval s of \rcapacity\ at each node, 
thus $\Nnum \cdot \nsize = \expp{21}$ \bitval s is the system \rcapacity.
The amount of storage needed by the repairer to store its programs and state 
generously is at most something like $\vsz = \expp{13}$ \bitval s.
Generally, $\nsize >> \vsz >> \Nnum$. 
We assume $\nsize \ge \Nnum$ and $\vsz << \Nnum \cdot \nsize$
in our bounds with respect to growing $\Nnum$.

Practical values of $\beta$ range from $2/3$ (triplication) to $1/20$ and smaller.
In the example, $\dsize = (1-\beta) \cdot \Nnum \cdot \nsize \approx \expp{21}$ \bitval s.
In practice nodes fail in a few years, e.g., $1/\lambda = 3$ years.

For practical systems, \srcdata\ is generally  maintained 
at the granularity of {\em objects}, and erasure codes are 
used to generate redundant data for each object.  
When using a $(\nnum, \knum, \rnum)$ erasure code, 
each object is segmented into $k$ source fragments, 
an encoder generates $\rnum = \nnum - \knum$ repair fragments 
from the $\knum$ source fragments, 
and each of these $\nnum = \knum + \rnum$ fragments is stored at a different node.  
An erasure code is MDS (maximum distance separable) 
if the object can be recovered from any $\knum$ of the $\nnum$ fragments.

\subsection{\Tradsystem s}
\label{tradsystem sec}

Replication is an example of a trivial MDS erasure code, i.e., 
each fragment is a copy of the original object.  
For example, triplication can be thought of as using the simple $(3,1,2)$ erasure code, 
wherein the object can be recovered from any one of the three copies.  
Some practical distributed storage systems use triplication. 

Reed-Solomon codes \cite{CCauchy95}, \cite{CRizzo97}, 
\cite{RFC5510} are MDS codes
that are used in a variety of applications and are a popular choice for storage systems.
For example, \cite{Huang12} and~\cite{Ford10} use a $(9,6,3)$ Reed-Solomon code, 
and~\cite{Dimakis13} uses a $(14,10,4)$ Reed-Solomon code.  These are
examples of {\em \tradsystem s}, i.e., systems that use small values 
of $\nnum$, $\knum$ and $\rnum$.

There are some issues that complicate the design of \tradsystem s.
For example, the data for each object is spread over a tiny fraction of the nodes,
i.e., in a system of $100,000$ nodes, 
triplication spreads the data for each object over only $3$ nodes, 
and a $(14,10,4)$ Reed-Solomon code spreads the data for each object over only $14$ nodes.
Thus, an issue for a \tradsystem\ is how to distribute the data for all the objects smoothly
over all the nodes.  

A typical approach is to assign each object to a 
{\em \placegroup}, where each \placegroup\ maps
to $\nnum$ of the $\Nnum$ nodes,  which  determines where 
the $n$ fragments of the object are stored. 
An equal amount of object data should be assigned to each \placegroup, 
and an equal number of \placegroup s should map a fragment to each node.
For \tradsystem s, Ceph~\cite{Ceph} recommends 
$\frac{100 \cdot \Nnum}{\nnum}$ \placegroup s,
i.e., 100 \placegroup s map a fragment to each node. 
A \placegroup\ should avoid mapping fragments to nodes with
correlated failures, e.g., to the same rack.
Pairs of \placegroup s should avoid mapping fragments to the same pair of nodes.
\Placegroup s are continually remapped as nodes fail and are added. 
These and other issues make the design of \tradsystem s challenging.

Since a small number $\rnum+1$ of \nfail s can cause \srcdata\ loss for \tradsystem s,
{\em \tradrepair} is used,  i.e., the repairer operates as quickly as practical to 
regenerate fragments lost from a node that permanently fails before another node fails,
and typically reads $\knum$ fragments to regenerate each lost fragment.
Thus, the peak \rrepairrate\ is higher than the average \rrepairrate, 
and the average \rrepairrate\ is $\knum$ times the \nfail\ \erasurerate. 

As highlighted in~\cite{Dimakis13}, the \rrepairrate\ needed 
to maintain \srcdata\ recoverability for \tradsystem s can be substantial.
Modifications of standard erasure codes have been  
designed for storage systems to reduce this rate,  
e.g., {\em local reconstruction codes} \cite{Gopalan12}, \cite{Dimakis13}, 
and {\em regenerating codes} \cite{Dimakis07}, \cite{Dimakis10}.  
Some versions of local reconstruction codes have been used in deployments, e.g., by Microsoft Azure.

\subsection{\Liqsystem s}

Another approach introduced in~\cite{Luby19} is {\em \liqsystem s},
which use erasure codes with large values of $\nnum$, $\knum$ and $\rnum$.  
For example, $\nnum =\Nnum$ and a fragment is assigned to each node for each object,
i.e., only one \placegroup\ is used for all objects.
The RaptorQ code \cite{CRaptorQ11},~\cite{RFC6330} is an example of 
an erasure code that is suitable for a \liqsystem, 
since objects with large numbers of fragments can be encoded and decoded
efficiently in linear time.
  
Typically $\rnum$ is large for a \liqsystem, thus \srcdata\
is unrecoverable only when a large number of nodes fail.
A \liqrepairer\ is lazy, i.e., repair operates to 
slowly regenerate fragments erased from nodes that have permanently failed.
The repairer reads $\knum$ fragments for each object to regenerate
around $\rnum$ fragments erased over time due to \nfail s, 
and the peak \rrepairrate\ is close to the average \rrepairrate.
The peak \rrepairrate\ for the \liqrepairer\ 
described in~\cite{Luby19} is within a factor of two 
of the lower bounds on the \rrepairrate,
and the peak \rrepairrate\ for the \advliqrepairer\ described 
in~\cite{LubyRich19} asymptotically 
approaches the lower bounds.

\section{Related work}
\label{related work sec}

The groundbreaking research of Dimakis et. al., described in~\cite{Dimakis07} 
and~\cite{Dimakis10}, is closest to our work:
An object-based distributed storage framework is introduced, 
and optimal tradeoffs between \storeoverhead\ 
and \lrepairer\ \rrepairrate\ are proved
with respect to repairing an individual object. 
\cite{Dimakis07} and~\cite{Dimakis10} describe
a framework, the types of repairers that fit into the framework,
and lower bounds on these types of repairers within the framework,
which are hereafter referred to as the {\em \Regframe},  
{\em \Regrepairer s}, and {\em \Reglower s}, respectively. 

The \Regframe\ was originally introduced to model 
repair of a single lost fragment, and is applicable to \tradrepair\ of a single object. 
The \Regframe\ is parameterized by $(n, k, d, \alpha, \gamma)$: 
$n$ is the number of fragments for the object (each stored at a different node); 
$k$ is the number of fragments from which the object must be recoverable; 
$d$ is the number of fragments used to generate a lost fragment at a new node 
when a node fails; $\alpha$ is the fragment size;
and $\gamma$ is the total amount of data generated and read across the
network to generate a fragment at a new node, i.e., 
$\gamma/d$ is the amount of data generated from 
each of $d$ fragments needed to generate a fragment at a new node.
\Reglower s on the \lrepairer\ read data rate prove necessary
 conditions on the \Regframe\ parameters used by any \Regrepairer\ 
to ensure that an individual object remains recoverable when using \tradrepair.

The \Reglower s were not originally designed to 
provide general lower bounds for a large system of nodes.
Nevertheless, it is interesting to interpret the \Regframe\ parameterized
by $(n, k, d, \alpha, \gamma)$ in the context of a system 
so that the \Reglower s can be as closely compared as possible to
the system level lower bounds proved in this paper.
Let $\dsize$ be the amount of \srcdata\ to be stored in the system.
Then, $n$ is set to the number of nodes $\Nnum$ in the system,
and $\alpha$ is set to the storage capacity $\nsize$ of each node, 
and thus the \storeoverhead\ $\beta$ is as shown in Equation~\eqref{beta eq}.

Since we want the best tradeoff possible between the amount of data
$\gamma$ read by the \Regrepairer\ to replace each failed node and the
\storeoverhead\ $\beta$, we set $k= d =\Nnum-1$. 
(At a general point in time a failed node is being replaced and
 there are only $\Nnum-1$ available nodes.)  
Thus, at the system level we consider the \Regframe\ with parameters 
\[(n=\Nnum, k=\Nnum-1, d=\Nnum-1, \alpha=\nsize, \gamma).\]

The \Regframe\ uses labeled acyclic directed graphs, where each directed
edge is labeled with the amount of data transferred from the node
at the tail to the node at the head of the edge, to represent
the actions of \Regrepairer s, and it is the properties of
these graphs that are used to prove the \Reglower s.
The labeled acyclic graphs restrict
the possible actions taken by \Regrepairer s as follows.
Suppose a node with \identifier\ $\idvar$ fails at time $\tvar$ 
and the next \nfail\ is at time $\tvar' > \tvar$.
A \Regrepairer\ is restricted to the following actions between 
time $\tvar$ and $\tvar'$:
\begin{itemize}
\item
For each node $\idvar'$ other than the failed node $\idvar$,
the \Regrepairer\ computes a function of the $\nsize$ \bitval s stored
at node $\idvar'$ to generate $\frac{\gamma}{\Nnum-1}$ \bitval s, 
and transfers the $\frac{\gamma}{\Nnum-1}$ \bitval s 
to the replacement node for $\idvar$.
\item
From the $\gamma$ \bitval s received at the replacement node for $\idvar$ 
from the $\Nnum-1$ nodes other than node $\idvar$, the \Regrepairer\ computes
a function of the $\gamma$ \bitval s to generate the $\nsize$ \bitval s
to be stored at the replacement node for $\idvar$.   
\end{itemize}
Thus, between time $\tvar$ and $\tvar'$, a fixed amount of data is transferred 
to the replacement node for $\idvar$ and no data is transferred to any other node;
once another node fails at time $\tvar'$, no more data is transferred 
to the replacement node for $\idvar$ until it fails again and 
is replaced with another replacement node; 
an equal amount of data $\frac{\gamma}{\Nnum-1}$ is read
and transferred from each of the $\Nnum-1$ non-failing node 
to the replacement node for $\idvar$.

\vspace{0.1in}
\begin{dimlemma}
\label{dim lemma}
The following holds as $\Nnum$ goes to infinity. 
For any \Regrepairer\ parameterized by 
$(\Nnum, \Nnum-1, \Nnum-1, \nsize, \gamma)$, if 
\begin{equation}
\label{dimakis lb eq} 
\gamma < \frac{\nsize}{2 \cdot \beta}.
\end{equation}
then the \srcdata\ cannot be reliably recovered at the end 
of any \nfseq\ with $\Nnum-1$ distinct \nfail s. 

\end{dimlemma}

\begin{proof}
Inequality (16) of \cite{Dimakis10} implies that if
\begin{equation}
\label{dimlb eq}
 \sum_{i=0}^{k-1} 
 \min\left\{\frac{(d-i) \cdot \gamma}{d}, \alpha\right\} < \dsize 
\end{equation}
for a \Regrepairer\ then the \srcdata\ cannot be reliably
recovered at the end of any \nfseq\ with $k$ distinct \nfail s.
With $k=d=\Nnum-1$, $\alpha = \nsize$, and using Equation~\eqref{beta eq}, 
we can rewrite Inequality~\eqref{dimlb eq} as 
\[
\sum_{j=0}^{\Nnum-1}
\min\left\{\frac{ j \cdot\gamma}{\Nnum}, \nsize\right\} 
< (1-\beta) \cdot \Nnum \cdot \nsize.
\]
As $\Nnum$ goes to infinity, we can approximate the sum by integration to yield:
\[
\int_{s=0}^{\frac{\nsize}{\gamma}} s \cdot \gamma \mbox{ {\rm d}s} + 
\int_{s=\frac{\nsize}{\gamma}}^1 \nsize \mbox{ {\rm d}s} <
 (1-\beta)\cdot \nsize .
\]
Simplifying yields Inequality~\eqref{dimakis lb eq}.
\qed
\end{proof}
\Dimlemma\ is tight, i.e.,
\cite{Dimakis07}, \cite{Dimakis10} describe \Regrepairer s 
with $\gamma = \frac{\nsize}{2 \cdot \beta}$ that 
maintain \srcdata\ recoverability for periodic \nfseq s.

\Dimlemma\ holds for any \Regrepairer\ and for any \nfseq\
with $\Nnum-1$ distinct \nfail s, even if the \Regrepairer\
is provided the \nfail s in advance.
However, the following repairer maintains recoverability of the \srcdata,
uses storage overhead $\beta = \frac{1}{\Nnum}$, 
and reads $\nsize$ \bitval s per \nfail\ for any \nfseq\
that is provided in advance.
The \srcdata\ of size $\dsize = (\Nnum-1)  \cdot \nsize$ 
is stored on $\Nnum-1$ nodes, and the remaining node is empty.
Just before the next \nfail, the repairer copies all data 
from the node that is going to fail to the empty
node, and the new node replacing 
the failed node becomes the empty node.

This repairer: (a) is not a \Regrepairer; (b) violates Inequality~\eqref{dimakis lb eq}
of \Dimlemma\ and yet the \srcdata\ can always be reliably recovered; 
(c) shows there is no non-trivial general lower bound
if the \nfseq\ is provided to the repairer in advance.
This shows that \Dimlemma\ is not a lower bound
that applies to all repairers, and that it is impossible to
prove non-trivial general lower bounds if the \nfseq\ is provided
to the repairer in advance.

The righthand side of Inequality~\eqref{dimakis lb eq} of 
\Dimlemma\  converges to the same value as the righthand side
of Inequality~\eqref{Poisson theorem eq} of \Poissonthm\
as $\Nnum$ goes to infinity $\beta$ goes to zero.
The main differences between \Dimlemma\ and \Poissonthm\
are the generality of the repairers to which the lower bounds apply
and the type of \nfseq s used to prove the lower bounds.  

\Dimlemma\ applies to \Regrepairer s within 
the restrictions of the \Regframe, i.e., 
a \Regrepairer\ that predictably reads a given amount
of data from each node and transfers a predictable amount of
data to a replacement node between \nfail s. 
On the other hand, \Poissonthm\ applies to any repairer, 
i.e., a repairer that can be completely unpredictable.
  
For a \Regrepairer\ that doesn't read enough data,
the \nfseq\ that causes the \srcdata\ 
to be unrecoverable in \Dimlemma\ is an atypical
\nfseq\ with $\Nnum-1$ distinct \nfail s. 
On the other hand, for a general repairer that doesn't
read enough data, the \nfseq\ that causes the \srcdata\
to be unrecoverable in \Poissonthm\ is a typical random
\nfseq\ that is chosen independently of the repairer.

The following examples show that repairers for practical systems
do not belong to the class of \Regrepairer s for which \Dimlemma\
applies, and thus \Dimlemma\ is not a 
lower bound on repairers in general.

As can be inferred from Section~\ref{tradsystem sec}, 
a typical repairer for a \tradsystem\ reads data 
from only a fraction of the $\Nnum$ nodes 
to replace the data on a failed node.  
Thus, repairers for \tradsystem s are not \Regrepairer s
for which the \Reglower\ of \Dimlemma\ applies.
 
\Liqsystem s repairers \cite{Luby19}, \cite{LubyRich19} 
transfer data incrementally to a replacement node 
over a constant fraction of $\Nnum$ \nfail s 
after the node it replaces fails.
Thus, repairers for \liqsystem s are not \Regrepairer s
for which the \Reglower\ of \Dimlemma\ applies.

\section{System model}
\label{model sec}

We introduce a model of distributed storage which is inspired by 
properties inherent and common to systems described in Section~\ref{overview sec}.
This model captures some of the essential features of any distributed
storage system.  All lower bounds are proved with respect to this system model.

There are a number of possible strategies beyond those
outlined in Section~\ref{overview sec} that could be used 
to implement a distributed storage system.
One of our primary contributions is to provide fundamental lower bounds
on the \rrepairrate\ needed to maintain \srcdata\ recoverability for {\em any}
distributed storage system, current or future, 
for a given \storeoverhead\ and \nfail\ rate.
Appendix~\ref{real system sec} provides details on how 
the system model introduced in this section applies to real systems.

\subsection{Architecture}

\begin{figure}
\centering
\includegraphics[width=0.5\textwidth]{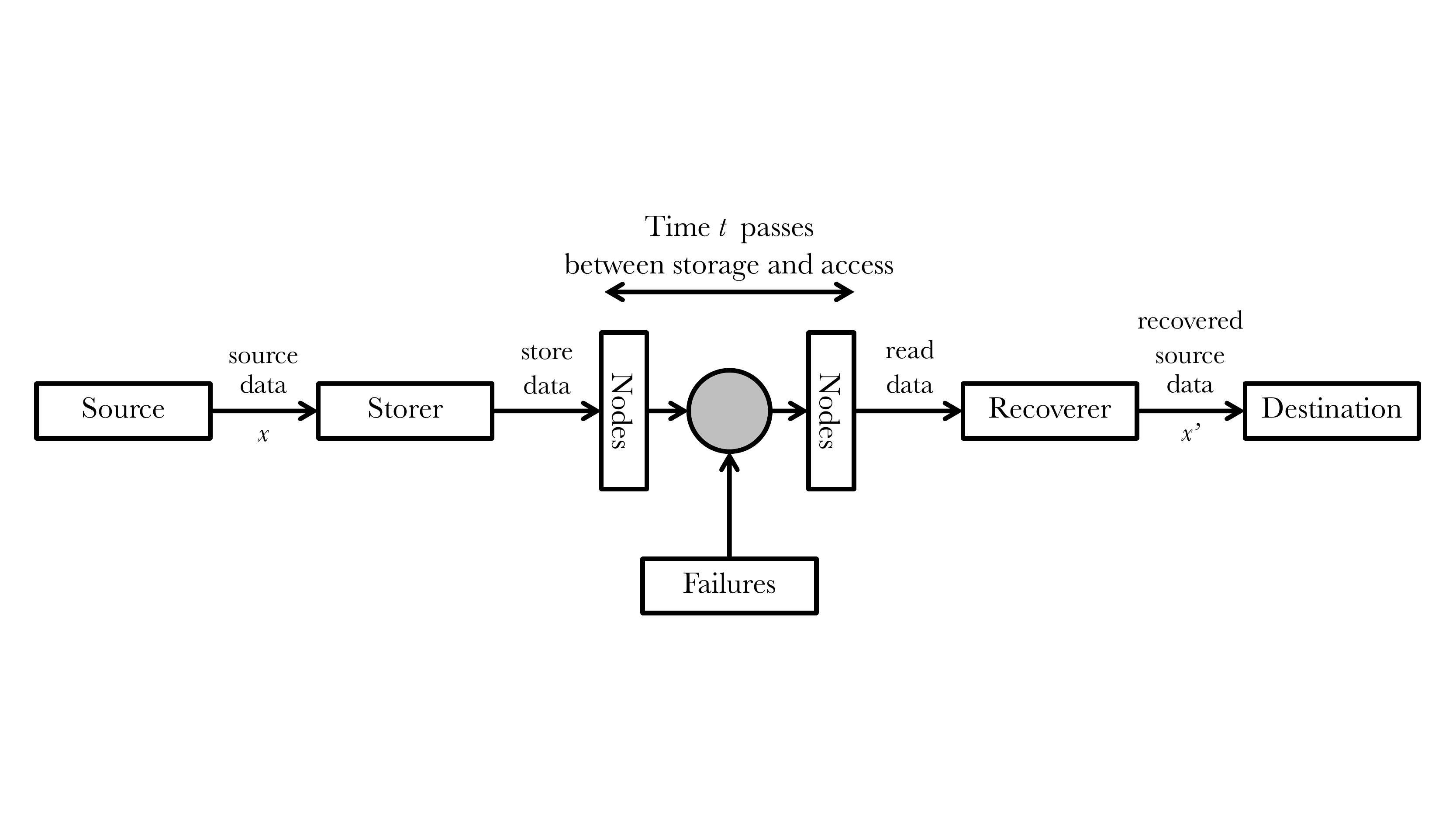}
\caption{Distributed storage architecture}
\label{storage_model fig}
\end{figure}

Figure~\ref{storage_model fig} shows an architectural overview 
of the distributed storage model.
A storer generates data from \srcdata\ 
$\xvar \in \{0,1\}^\dsize$ received from a source,
and stores the generated data at nodes.
In our model,  the \srcdata\ $\Xvar$ 
is randomly and uniformly chosen,
where $\Xvar \inu \{0,1\}^\dsize$ is a random variable and
$\inu$ indicates randomly and uniformly chosen.
Thus, $\len{\Xvar}= \entropy{\Xvar} =\dsize$, where
$\len{\Xvar}$ is the length of $\Xvar$ and $\entropy{\Xvar}$ 
is the entropy of $\Xvar$. 

\noindent

\begin{figure}
\centering
\includegraphics[width=0.50\textwidth]{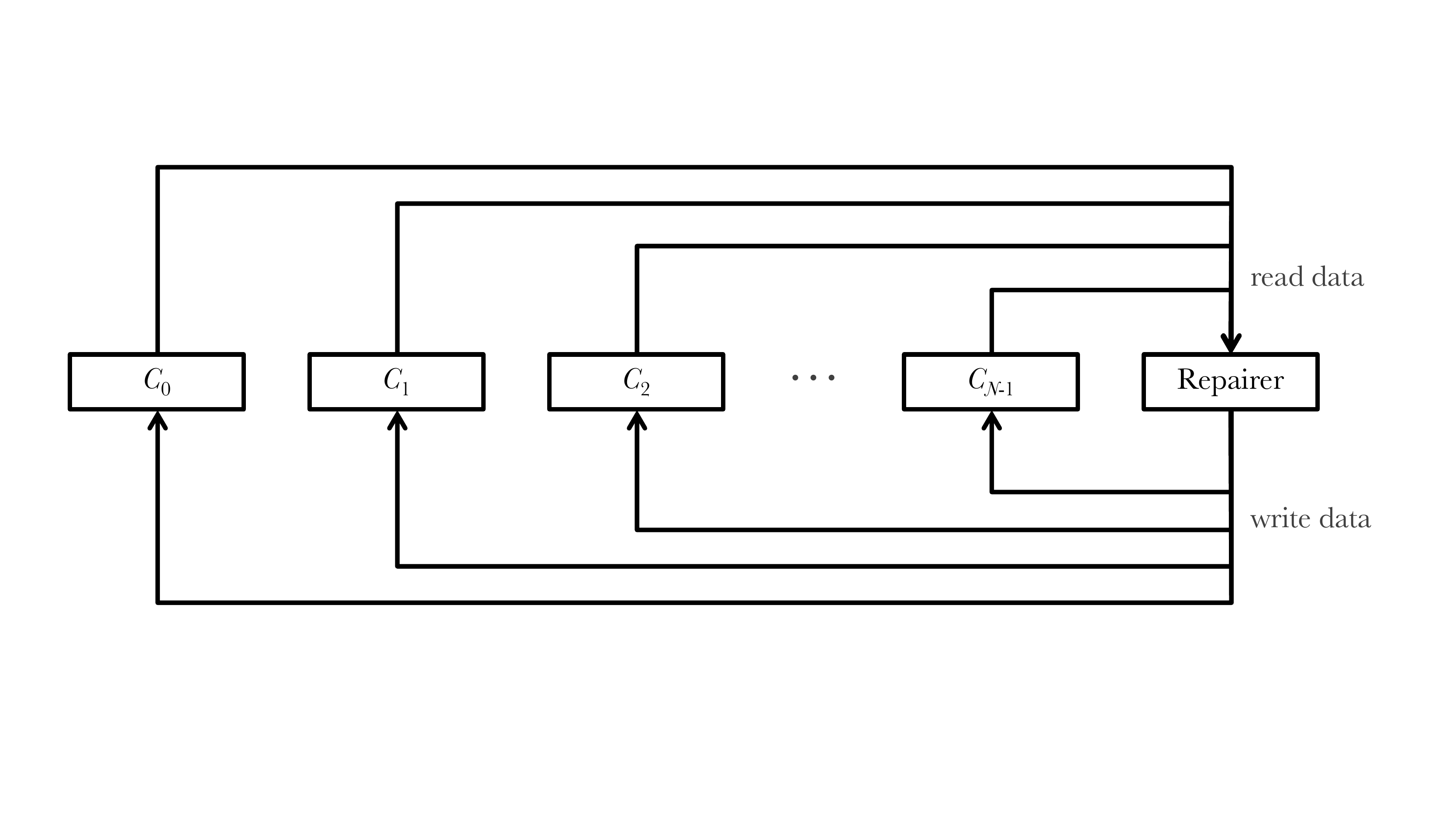}
\caption{Storage nodes and repairer model.}
\label{storage_nodes fig}
\end{figure}

Figure~\ref{storage_nodes fig} shows the nodes of the distributed storage system, 
together with the network that connects each node to a repairer.
Each of $\Nnum$ nodes 
$\Cdata_0,\ldots,\Cdata_{\Nnum-1}$ can store $\nsize$ \bitval s,
and the {\em \rcapacity} is $\Nnum \cdot \nsize$.

\noindent

As nodes fail and are replaced, a repairer continually reads data from the nodes, 
computes a function of the read data, and writes the computed data back to the nodes.  
The repairer tries to ensure that the \srcdata\ can be recovered at any time
from the data stored at the nodes.

As shown in Figure~\ref{storage_model fig},
after some amount of time $\tvar$ passes, a recoverer reads data from the nodes to
generate $\xvar'$, which is provided to a destination, where $\xvar$ 
is reliably recovered if $\xvar' = \xvar$.
The goal is to maximize the amount of time $\tvar$ the recoverer can reliably recover $\xvar$.

\subsection{\Nfail s}

A \nfseq\ determines when and what nodes fail as time passes.
A \nfseq\ is a combination of two sequences,
a {\em \timeseq}
 \[\tvar_{0} \le \tvar_{1} \le \cdots 
 \le\tvar_{i} \cdots, \]
where for {\em index} $i$,  $\tvar_{i}$ is the {\em time} at which a node fails, 
and an {\em \identifierseq}
\[\idvar_0,\idvar_{1},\ldots,\idvar_{i},\ldots,\]
where $\idvar_{i}$ is the {\em \identifier} of the node that fails at time $\tvar_{i}$.

All $\nsize$ \bitval s stored at node $\idvar_{i}$ are 
immediately erased at time $\tvar_{i}$ when the node fails, 
where erasing a bit means setting its value to zero.  
This can be viewed as immediately replacing 
a failed node with a replacement node 
with storage initialized to zeroes.
Thus, at each time there are $\Nnum$ nodes.

A primary objective of practical distributed storage architectures is to distribute the
components of the system so that failures are as independent as possible.
{\em \Ptdist s} are an idealization of this primary objective, 
and are often used to model and evaluate distributed storage systems in practice.
For a \Ptdist\ with rate $\lambda$, the time between when a node is 
initialized and when it fails is an independent 
exponential random variable with rate $\lambda$,
i.e., $\frac{1}{\lambda}$ is the average lifetime of a node between
when it is initialized and when it fails.
Our main lower bounds in Section~\ref{main sec} 
are with respect to \Ptdist s.

\subsection{Network}
\label{network sec}

The model assumes there is a network interface between 
each node and the repairer over which all data from and to the node travels.
One of the primary lower bound metrics is the amount of data
that travels over interfaces from nodes to the repairer, which is counted
as data read by the repairer.   For the lower bounds, this is the only
network traffic that is counted. All other data traffic within
the repairer, i.e. data traffic internal to a distributed repairer,
data traffic over an interface from the repairer to nodes,
or any other data traffic that does not travel over an interface 
from a node to the repairer, is not counted for the lower bounds.
It is assumed that the network is completely reliable, and that
all data that travels over an interface from a node to the repairer
is instantly available everywhere within the repairer.   

\subsection{Storer}
\label{storer sec}
A storer takes the \srcdata\ $\xvar$ and generates and
stores data at the nodes in a preprocessing step 
when the system is first initialized and before there are any \nfail s.
We assume that the recoverer can reliably recover $\xvar$ 
from the data stored at the nodes 
immediately after the preprocessing step finishes. 

For simplicity, we view the storer preprocessing step as part of the repairer, 
and any data read during the storer preprocessing step is not counted in the lower bounds.

For the lower bounds, there are no assumptions about 
how the storer generates the stored data from the \srcdata,
i.e. no assumptions about any type of coding used, no assumptions
about partitioning the \srcdata\ into objects, etc.
As an example, the \srcdata\ can be encrypted, compressed,
encoded using an error-correcting code or erasure code, replicated, 
or processed in any other way known or unknown to generate
the stored data, and still the lower bounds hold.  Analogous
remarks hold for the repairers described next.

\subsection{Repairer}
\label{repairer initial sec}
 
A repairer for a system is a process that operates as follows.
The \identifier\ $\idvar_{i}$ is provided to a repairer
at time $\tvar_{i}$, which alerts the repairer that all $\nsize$
\bitval s stored on node $\idvar_{i}$ are lost at that time.
As nodes fail and are replaced, the repairer reads data over interfaces from nodes, 
performs computations on the read data, and writes computed data
over interfaces to nodes.
A primary metric is the number of \bitval s the repairer reads over interfaces
from storage nodes.

Appendix~\ref{repairer sec} provides a detailed description of repairers,
including \lrepairer s.

\subsection{Recoverer}
\label{recovery sec}

For any repairer $\Rfnc$ there is a recoverer $\Afnc$ 
such that if the \srcdata\ is $\xvar$ and the state at time $\tvar$ is 
$\Sdata(\tvar)$ when the repairer is $\Rfnc$ 
then $\Afnc(\Sdata(\tvar))$ should be equal to $\xvar$.

\Srcdata\ $x$ is {\em recoverable} at time $\tvar$
with respect to repairer $\Rfnc$ and recoverer $\Afnc$ if
$\Afnc(\Sdata(\tvar)) = \xvar$.
\Srcdata\ $x$ is {\em unrecoverable} at time $\tvar$
with respect to repairer $\Rfnc$ and recoverer $\Afnc$ if
$\Afnc(\Sdata(\tvar)) \not= \xvar$.

\subsection{System State}
\label{system state sec}

At time $\tvar$,  let
$\Vdata(\tvar)$ be the \bitval s stored in the global memory of the repairer, 
where $\vsz = \len{\Vdata(\tvar)}$,
\[ \Cdata_0(\tvar),\ldots,\Cdata_{\Nnum-1}(\tvar) \]
be the \bitval s stored at nodes $0,\ldots,\Nnum-1$,
respectively, where $\nsize = \len{\Cdata_j(\tvar)}$ is the 
capacity of each node $j$, and
\[ \Sdata(\tvar) = \{\Vdata(\tvar), 
\{ \Cdata_0(\tvar),\ldots,\Cdata_{\Nnum-1}(\tvar) \} \} \]
is the global state of the system at time $\tvar$, 
where $\ssize = \len{\Sdata(\tvar)}$.
Thus, 
\[  \ssize = \vsz + \Nnum \cdot \nsize \]
is the size of the global system state at any time $\tvar$.

\section{Guide for lower bound proofs}
\label{lbover sec}

The ultimate goal is to prove that Inequality~\eqref{rrate lb eq}
holds with respect to the \Ptdist\ for any repairer.

In practice, the timing of \nfail s, and the identity of which
nodes fail, are not known in advance, and thus
repairers must handle these uncertainties.
A much simpler model for repairers to handle is a 
{\em periodic} \nfseq, i.e., 
the time between consecutive \nfail s  is a constant known to the repairer.
Many of the lower bounds we prove hold
for periodic \nfseq s, and the only uncertainty is which nodes fail. 

Lower bounds for $\beta \ge \frac{1}{2}$ are not of great interest,
since for $\beta =  \frac{1}{2}$, the repairer that maintains
a duplicate copy of the \srcdata\ succeeds in maintaining recoverability
of the \srcdata\ forever for periodic \nfseq s.
Furthermore, in practice, the interest is to decrease $\beta$ as much
as possible, and thus we hereafter restrict attention to $\beta \le \frac{1}{2}$. 

Section~\ref{universal sec} introduces the notion of a {\em phase},
where a phase is a \nfseq\ of a specified number $\Mdata$ of distinct \nfail s.
Let $\rrate_i$ be the average \rrepairrate\ of a repairer 
over the first $i$ \nfail s of the phase for a periodic \nfseq.
The overall idea of the lower bound proof is to show that, 
for any repairer, either there is an $i \le \Mdata$ such that
$\rrate_i$ is above a lower bound rate, 
or at the end of the phase the \srcdata\ is unrecoverable.

Section~\ref{universal sec} shows that the system state at the
end of the phase can be generated from $\Ddata$, 
where $\Ddata$ is the concatenation of the data read 
from nodes that fail in a phase before they fail 
and the data at nodes at the start of the phase
that don't fail in the phase.
The crucial but simple \Complemma\ and \Compcorollary\ show that
if $\len{\Ddata} < \dsize$ then the \srcdata\ is unrecoverable from $\Ddata$,
where $\len{\Ddata}$ is the length of $\Ddata$.
Since the system state at the end of the phase can be generated
from $\Ddata$, this implies that if $\len{\Ddata} < \dsize$
then the \srcdata\  is unrecoverable
from the system state at the end of the phase.

Section~\ref{erepairer sec} introduces a restricted class of repairers,
\Erepairer s, that predictably read an equal amount of
data from each node between \nfail s.  \Erepairer s are introduced
for two reasons: (1) they are similar to (but more general than) the
\Regrepairer s discussed in Section~\ref{related work sec}; 
(2) based on the framework introduced in Section~\ref{universal sec}, 
the lower bound proofs for \Erepairer s are easy and straightforward.
\Elemma\ of Section~\ref{erepairer sec} shows that if the
predictable read rate of the \Erepairer\ is below a lower
bound then the \srcdata\ is unrecoverable with very high
probability at the end of the phase.

The \Elemma\ lower bound for \Erepairer s holds for any \nfseq\
with distinct \nfail s, even if the \nfseq\ is known in advance to the \Erepairer.
However, as outlined in Section~\ref{related work sec}, 
there is no non-trivial lower bound for general repairers 
if the \nfseq\ is known in advance.
Thus, using random \nfseq s for which the repairer cannot predict 
which future nodes will fail is key to proving general lower bounds.

Repairer actions can be unpredictable: 
A repairer may read different amounts of data from 
individual nodes between \nfail s, 
and may read different amounts of data in aggregate from
all nodes between different \nfail s.
The repairer actions can depend on the \srcdata,
which nodes have failed in the past, and the timing of past \nfail s.

Section~\ref{core lower sec} provides the technical core 
of the lower bound proofs that use random \nfseq s with
distinct \nfail s to prove lower bounds on general repairers
that may act unpredictably.  
The proof of \Corelemma\ in Section~\ref{core lower sec} 
is the most technically challenging proof in the paper.  
It shows that, for any repairer, when the \identifierseq\  
consists of randomly chosen distinct \identifier s
within a phase of $\Mdata$ \nfail s
then there is only a tiny probability $\delcore$ that
$\len{\Ddata} \ge \dsize - \nsize$ when $\rrate_i$
is below a lower bound for each $i \le \Mdata$.
\Supertheorem\ is used to prove \Corelemma, and may
 be of independent interest.
\Coretheorem\ directly uses \Corelemma\ to show that
with very high probability there is either an $i \le \Mdata$
where the \rrepairrate\ $\rrate_i$ for the repairer 
up to the $i^{\rm th}$ \nfail\ in the phase 
is above a lower bound or else 
the \srcdata\ is unrecoverable at the end of the phase.

A phase terminates early if the repairer reads enough
data from all nodes in a prefix of the phase, 
i.e., a phase is terminated at the 
first index $i \le \Mdata$ where $\rrate_i$ is above a lower bound,
and another phase is started at that point.  
The overall lower bounds are proved by stitching 
together consecutive phases.  
Thus, the lower bound holds over the \nfseq\ within each 
stitched together phase.
There are some technical issues with stitching together phases.
The actions of the repairer have an influence on when
one phase ends and the next phase begins.  Since the distinct \nfail s
within a phase depend on when the phase starts, the repairer
has an influence on the \nfseq.  

What we would like instead are lower bounds where the \nfseq\
is chosen completely independently of the repairer,
which is achieved in Section~\ref{main sec}.
\Geolemma\ together with \Unitheorem\ shows that the lower bounds
for periodic \nfail s shown in \Coretheorem\ still apply when the
\nfail s within a \nfseq\ are chosen uniformly at random and are
no longer required to be distinct within a phase.
\Poissonthm, the main result of this research, 
shows that the lower bounds of \Unitheorem\ 
extend when the \timeseq\ of the \nfseq\ is Poisson 
distributed instead of being restricted to being periodic.
Thus, \Poissonthm\ shows that for any repairer
the lower bounds apply when nodes fail independently according
to a Poisson process.   

\section{Emulating repairers in phases}
\label{universal sec}

We prove lower bounds based on considering 
the actions of a repairer $\Rfnc$, or \lrepairer\ $\Rfnc$,
running in phases.  Each phase considers
a \nfseq\ with $\Mdata$ \nfail s, where each
of the $\Mdata$ \nfail s within a phase are distinct, as described in more detail below.

For any $\Mdata \le \Nnum$, we write 
\[ \ang{\idvar_0,\ldots,\idvar_{\Mdata-1}} \]
when all  $\Mdata$ \identifier s are distinct, i.e., $\idvar_{i} \not= \idvar_{i'}$
for $0 \le i \not= i' \le \Mdata-1$, which we hereafter refer to as a {\em \didentifierseq}.
We write 
\[ \ang{\idvar_0,\ldots,\idvar_{j-1}, \IDvar_j,\ldots,\IDvar_{\Mdata-1}} \]
when $\ang{\idvar_0,\ldots,\idvar_{j-1}}$ are distinct \identifier s,
random variable $\IDvar_j$ is defined as
\[ \IDvar_{j} \inu \{0,\ldots,\Nnum-1\} - \{\idvar_0,\ldots,\idvar_{j-1}\}, \]
and for $i =j+1,\ldots,\Mdata-1$, random variable $\IDvar_i$ is defined as
\[ \IDvar_{i} \inu \{0,\ldots,\Nnum-1\} - \{\idvar_0,\ldots,\idvar_{j-1},\IDvar_j,\ldots\IDvar_{i-1}\}, \]
where $\inu$ indicates randomly and uniformly chosen.
Thus, $\ang{\idvar_0,\ldots,\idvar_{j-1}, \IDvar_j,\ldots,\IDvar_{\Mdata-1}}$ is a
distribution on \didentifierseq s.

A phase consists of executing $\Rfnc$ on 
a \nfseq\ $(\tseq,\idseq)$, where 
\[ \tseq = \{\tvar_{0}, \tvar_{1} , \ldots ,\tvar_{\Mdata-1} \} \]
is the \timeseq\ and
\[ \idseq = \ang{\idvar_0, \idvar_{1}, \ldots, \idvar_{\Mdata-1}} \]
is the \didentifierseq\ that is revealed to $\Rfnc$ 
as the phase progresses.

For $i \in \{0,\ldots,\Mdata-1\}$, let 
\[
\tseq_i = \{\tvar_0,\ldots,\tvar_{i}\}
\] 
be a prefix of $\tseq$, and let 
\[
 \idseq_i = \ang{\idvar_0,\ldots,\idvar_{i}}
 \] 
be a prefix of $\idseq$. 
 
Fix repairer $\Rfnc$, recoverer $\Afnc$,
\timeseq\ $\tseq$ and \identifierseq\ $\idseq$, and $\xvar$.
The variables defined below depend on these parameters,
but to simplify notation this dependence is not explicitly expressed 
in the variable names.

For
$i \in \{0,\ldots,\Mdata-1\}$, $j \in \{0,\ldots,\Nnum-1\}$, 
let $\Rdata_{i,j}$
be the data read by $\Rfnc$ from node $j$ 
between $\tvar_{0}$ and $\tvar_i $
with respect to $\xvar$, $\Rfnc$, $(\tseq_i,\idseq_{i-1}))$, 
and let $\rsize_{i,j} = \len{\Rdata_{i,j}}$
be the size (or length) of $\Rdata_{i,j}$.
(If $\Rfnc$ is a \lrepairer, then $\Rdata_{i,j}$ is the 
the locally-computed \bitval s read by $\Rfnc$ over the interface
from node $j$ between $\tvar_{0}$ and $\tvar_i $.)
For $i \in \{0,\ldots,\Mdata-1\}$, let $\Rdata_i$ be the data read
from all nodes by $\Rfnc$ between $\tvar_{0}$ and $\tvar_i $, 
and let
\[ \rsize_{i} = \sum_{j \in \{0,\ldots,\Nnum-1\}} 
\rsize_{i,j} = \len{\Rdata_i} \]
be the total amount of data read from all nodes by $\Rfnc$
between $\tvar_{0}$ and $\tvar_i $.
 
Let 
\[
\rfsize_i = \len{\Rdata_{i,\idvar_{i}}},
\]
where $\Rdata_{i,\idvar_{i}}$
is the data read from the node $\idvar_i$ between $\tvar_{0}$ and 
the time $\tvar_{i}$ of its failure 
with respect to $\xvar$, $\Rfnc$, $(\tseq_i,\idseq_i)$.
Let
\begin{equation*}
\rfsize = \sum_{i=0}^{\Mdata-1} \rfsize_{i}
\end{equation*}
be the total length of data read by $\Rfnc$ in the phase
from nodes that fail before their failure in the phase.

Before a phase begins, the storer generated and stored data at the nodes
based on \srcdata\ $\xvar$, and the repairer $\Rfnc$ has been
executed with respect to a \nfseq\ up till time $\tvarm_0$, 
where $\tvarm_0$ is just before 
the time of the first \nfail\ of the phase at time $\tvar_{0}$.  
We assume that the recoverer $\Afnc$ can recover \srcdata\ $\xvar$ 
from the state $\Sdata(\tvarm_0)$.

\subsection{Compressed state $\Ddata$}
\label{compress defn sec}

For this subsection,  we fix repairer $\Rfnc$, recoverer $\Afnc$,
\timeseq\ $\tseq$ and \identifierseq\ $\idseq$ and \srcdata\ $\xvar$.
The variables defined below depend on these parameters,
but to simplify notation this dependence is not explicitly expressed 
in the variable names.

We conceptually define two executions 
of a phase with respect to $\xvar$, $\Rfnc$, $\Afnc$, and $(\tseq, \idseq)$.
The first execution runs $\Rfnc$ normally from $\tvarm_0$ to $\tvarp_{\Mdata-1}$
starting system state $\Sdata(\tvarm_0)$ and ending in $\Sdata(\tvarp_{\Mdata-1})$,
where $\tvarm_0$ is just before $\tvar_{0}$, 
and $\tvarp_{\Mdata-1}$ is just after $\tvar_{\Mdata-1}$.
Thus, the \nfail s at times $\tvar_{0}$ and $\tvar_{\Mdata-1}$
are within the phase, but $\Rfnc$ does not read 
any \bitval s before $\tvar_{0}$ or after $\tvar_{\Mdata-1}$ in the phase.

Let $\RFdata$ be the concatenation of \bitval s read by $\Rfnc$
from nodes that fail before they fail in the phase,  concatenated in
the order they are read.  Thus $\RFdata$ contains all the \bitval s of
\begin{equation*}
\{ \Rdata_{i,\idvar_{i}} : i \in \{0,\ldots,\Mdata-1\} \}
\end{equation*}
and $\rfsize = \len{\RFdata}$,
but the order of the \bitval s in $\RFdata$ is 
defined by the order in which they are read by $\Rfnc$.  

Let
\begin{equation*}
\Ddata = \{ \Vdata(\tvarm_0), 
\{ \Cdata_j(\tvarm_0): j \not\in \idseq\}, \RFdata\},
\end{equation*}
which we hereafter refer to as the {\em compressed state} 
with respect to $\xvar$, $\Rfnc$, and $(\tseq, \idseq)$,
and thus
\begin{equation}
\label{sddata eq}
\len{\Ddata} = \vsz + (\Nnum - \Mdata) \cdot \nsize+ \rfsize.
\end{equation}

The second execution is an exact replay of the first execution, 
i.e., the repairer $\Rfnc$ reads, computes, and writes exactly
the same \bitval s at the same times as in the first execution with respect to
the \nfseq\ $(\tseq,\idseq)$ to arrive in the same final state 
$\Sdata(\tvarp_{\Mdata-1})$ as the first execution. 
However, the second execution uses the compressed state $\Ddata$ 
in place of $\Sdata(\tvarm_0)$ as the starting point of the execution.
The initial global memory state of $\Rfnc$ is set to $\Vdata(\tvarm_0)$
at time $\tvarm_0$. For all $j \notin \idseq$, the state of node $j$ is initialized to 
$\Cdata_j(\tvarm_0)$ at  time $\tvarm_0$.

Initially at time $\tvarm_0$, $f:\{0,\ldots,\Nnum-1\} \rightarrow \{0,1\}$
is set as:
\begin{align*}
f(j) = 0  & \mbox{ for all }   j \not\in \idseq \\
f(j) = 1 & \mbox{ for all }   j \in \idseq.
\end{align*}
Let $\tvar$ be a time within the phase, i.e., 
$\tvarm_0 \le \tvar \le \tvarp_{\Mdata-1}$.

Suppose at time $\tvar$ that $\Rfnc$ is to read \bitval s 
over the interface from node $j$:  if $f(j) == 0$ then
the requested \bitval s are read from $\Cdata_j(\tvar)$ exactly
the same as in the first execution; 
if $f(j) == 1$ then the requested \bitval s are provided to $\Rfnc$ 
from the next consecutive portion of $\RFdata$ not yet provided to $\Rfnc$,
which, by the properties of $\RFdata$,
are guaranteed to be the \bitval s read from node $j$
 at time $\tvar$ in the first execution. 

Suppose at time $\tvar$ that $\Rfnc$ is to write \bitval s to node $j$:
if $f(j) == 0$ then the \bitval s are written to $\Cdata_j(\tvar)$
exactly the same as in the first execution;
if $f(j) == 1$ then the write is skipped since whatever \bitval s are subsequently
read from node $j$ up till the time node $j$ fails are already part of $\RFdata$. 

For each $i \in \{0,\ldots, \Mdata-1 \}$, 
$f(\idvar_i)$ is reset to $0$ and the state of node $\idvar_i$, 
$\Cdata_{\idvar_i}(\tvar_i)$,  is initialized to zeroes
at time $\tvar_i$.

If $\Rfnc$ is a \lrepairer\ instead of a repairer then
when $\Rfnc$ is to produce and read locally-computed \bitval s 
over the interface from node $j$ at time $\tvar$ and $f(j) == 0$ 
the requested \bitval s are locally-computed
by $\Rfnc$ based also on $\Cdata_j(\tvar)$. 

It can be verified that the state of the system is 
$\Sdata(\tvarp_{\Mdata-1})$ at the end of the second execution,
whether $\Rfnc$ is a repairer or a \lrepairer.
Thus, $(\Sdata(\tvarp_{\Mdata-1}),(\tseq,\idseq))$ can be generated from
$(\Ddata,(\tseq,\idseq))$ based on $\Rfnc$.

\subsection{Viewing $\Ddata$ as a cut in an acyclic graph}
\label{acyclic sec}

\begin{figure}
\centering
\includegraphics[width=0.5\textwidth]{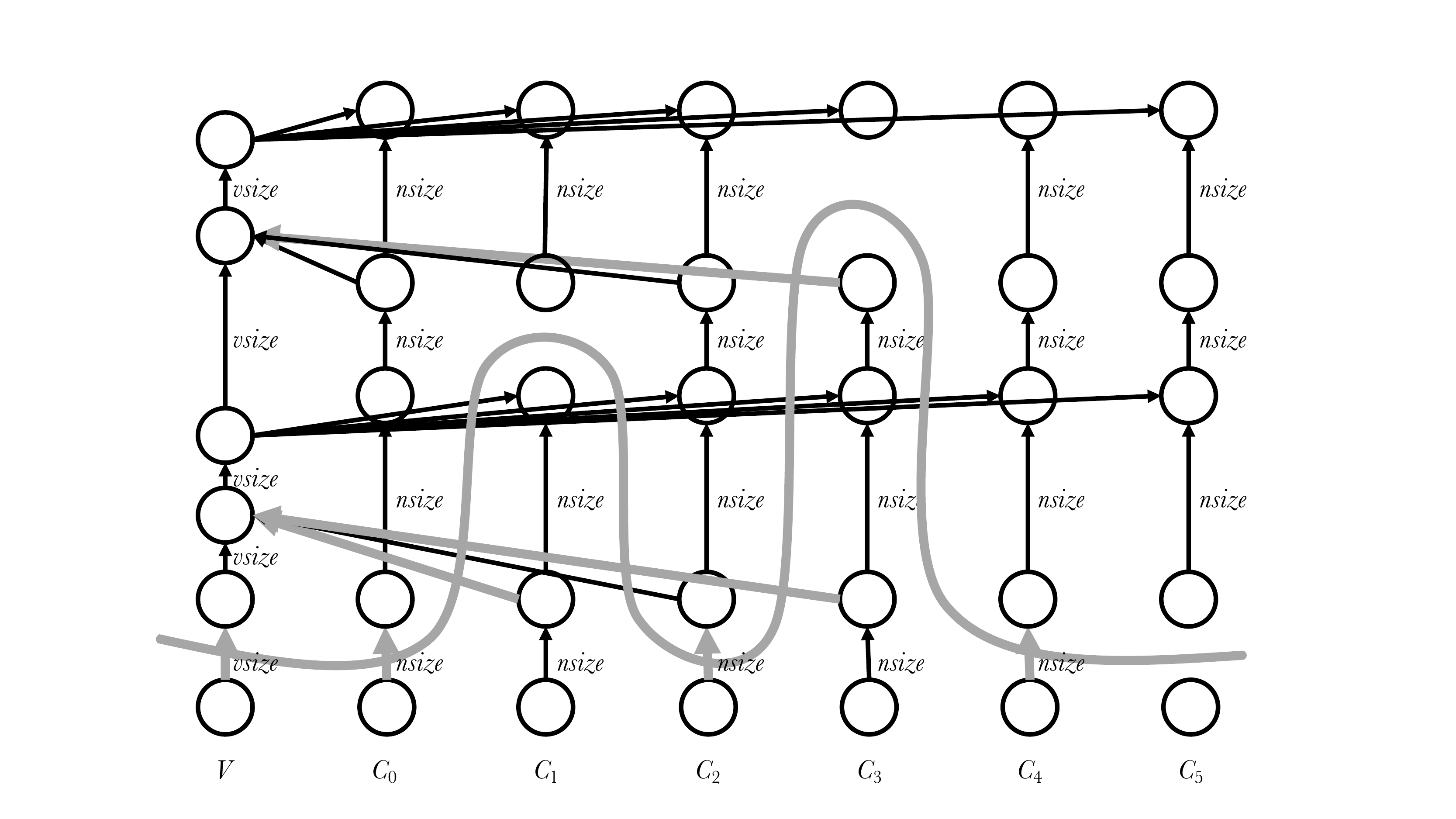}
\caption{Viewing $\Ddata$ as a cut in an acyclic graph}
\label{acyclic_graph fig}
\end{figure}

Similar to \cite{Dimakis07}, \cite{Dimakis10},
the compressed state $\Ddata$ can be viewed as a cut in an
acyclic graph.  An example of the acyclic graph is shown 
in Figure~\ref{acyclic_graph fig}, where there are $\Nnum = 6$ storage nodes.  
The beginning of the phase is at the bottom, 
and going vertically up corresponds to time flowing forward.
The leftmost vertical column is for the global memory $\Vdata$ 
of size $\vsz$, and there is a vertical column for each of 
the $\Nnum$ storage nodes, each of size $\nsize$.
The bottom row of vertices corresponds to the system state $\Sdata(\tvarm_0)$
at the start of the phase; the second from the bottom row of vertices corresponds
to the system state $\Sdata(\tvarp_0)$;
the top row of vertices corresponds to the 
system state $\Sdata(\tvarp_{\Mdata-1})$ at the end of the phase.
The vertices in a storage node column correspond to the state of that 
storage node over time, where edges flowing out of the column
correspond to data transfer out of the node, and edges flowing 
into the column correspond to data transfer into the node.  
Similar remarks hold for the global memory column.

The edges pointing vertically up are labeled with
the capacity of the corresponding entity, i.e., $\vsz$ is the capacity of 
$\Vdata$, and $\nsize$ is the capacity of each storage node.
The non-vertical edges that connect a first vertex to a second vertex 
correspond to a data transfer, where the label of the edge corresponds 
to the amount of data transferred.

In the example shown in Figure~\ref{acyclic_graph fig}, 
$\Cdata_5$ fails at time $\tvar_0$, $\Cdata_1$ fails slightly later, 
and $\Cdata_3$ fails at a slightly later time.
Each node that fails is replaced with an empty node,
and thus there is no edge from a vertex corresponding to 
a node just before it fails to the vertex above corresponding to the 
replacement node. 
Thus, $\Ddata$ includes all the data transferred along the edges that emanate from
the vertices in the columns corresponding to $\Cdata_1$, $\Cdata_3$, and $\Cdata_5$
before their failures, where these edges are shown in gray in Figure~\ref{acyclic_graph fig}.

In the example shown in Figure~\ref{acyclic_graph fig}, 
$\Cdata_0, \Cdata_2$ and $\Cdata_4$ do not fail before the end of the phase. 
Thus, $\Ddata$ includes the $\nsize$ \bitval s of data transferred along 
the vertical edges from the bottom row to the second from the bottom row 
for each column corresponding to these storage nodes,
where these edges are shown in gray in Figure~\ref{acyclic_graph fig}.   
In addition, $\Ddata$ includes the $\vsz$ \bitval s of data transferred
along the vertical edge from the bottom row to the second from the bottom row
for the first column corresponding to global memory $\Vdata$. 

The cut corresponding to $\Ddata$ is shown in Figure~\ref{acyclic_graph fig}
as the curved gray line, where $\len{\Ddata}$ is the sum of the labels of the edges
crossing the cut from the vertices below the cut.

The \bitval\ values of $\Ddata$ determine the edges and 
the edge label values in the acyclic graph, i.e., the edges
and edge label values in the acyclic graph
depend on the \bitval\ values stored at the vertices in the graph. 
This is unlike the acyclic graph representation in 
\cite{Dimakis07}, \cite{Dimakis10}, where the edges and 
the edge label values are independent of the \bitval\ values stored 
at the vertices in the graph.

Although the acyclic graph visualization of $\Ddata$ 
provides some good intuition, Section~\ref{compress defn sec}
provides the formal definition of $\Ddata$ and its properties.
 
\noindent

\subsection{Compression lemma}
\label{compress lemma sec}

For this subsection,  we fix repairer $\Rfnc$, recoverer $\Afnc$,
\timeseq\ $\tseq$ and \identifierseq\ $\idseq$.
The variables defined below depend on these parameters,
but to simplify notation this dependence is not explicitly expressed 
in the variable names.

The value of the \srcdata\ $\xvar$ is a variable in this subsection. 
Random variable $\Xvar \inu \{0,1\}^{\dsize}$ 
is uniformly distributed on the \srcdata.
We let 
\[ [\xvar : \Ddata :\xpvar ] \] 
indicate that \srcdata\ $\xvar$ is mapped before the start of the phase to
a value of $\Sdata(\tvarm_0)$ by $\Rfnc$, which in turn is mapped by $(\tseq,\idseq)$
to a value $\Ddata$ by the first execution of the emulation of $\Rfnc$, 
which is mapped to a value of 
$\Sdata(\tvarp_{\Mdata-1})$ by $(\tseq,\idseq)$ by the second execution 
of the emulation of $\Rfnc$, which in turn is mapped by $\Afnc$ to $\xpvar$.

\vspace{0.1in}
\begin{complemma}
\label{compression lemma}
Fix any repairer or \lrepairer\ $\Rfnc$, recoverer $\Afnc$,
\timeseq\ $\tseq$, and \didentifierseq\  $\idseq$. Let $\ell \le \dsize$.
Then,
\begin{gather*}
\nonumber
\Probi{\Xvar}{[\Xvar : \Ddata :\xpvar ]  \mbox{ s.t. } \len{\Ddata} \le \dsize-\ell 
\wedge \Xvar = \xpvar} \le 2^{-\ell+1}. 
\end{gather*}
\end{complemma}
\begin{proof}
Fix $\Rfnc$, $\Afnc$, and $(\tseq,\idseq)$.
The size of the set 
\[ \left\{ \xpvar : \exists \xvar \mbox{ s.t. } [\xvar : \Ddata :\xpvar ]
\wedge \len{\Ddata} \le \dsize-\ell \right\}
\]
is at most $2^{\dsize-\ell+1}$ since there are at most $2^{\dsize-\ell+1}$
\bitval-strings of size at most $\dsize-\ell$ and 
 any fixed value of $\Ddata$ maps to a unique value $\xpvar$ in the second execution.
Thus, there are at most $2^{\dsize-\ell+1}$ values for $\xvar$
such that $\xvar = \xpvar$ when $\len{\Ddata} \le \dsize-\ell$. 
\qed
\end{proof}

\subsection{Compression corollary}
\label{compress corr sec}

Let
\begin{equation}
\label{sfdata eq}
\osize = \Nnum \cdot \nsize - \dsize + \vsz,
\end{equation}
and let 
\begin{equation}
\label{fdata eq}
\Fdata = \left\lceil \frac{\osize}{\nsize} \right\rceil
\end{equation}
be the minimal number of nodes so that 
$\Fdata \cdot \nsize \ge \osize$.
Let
\begin{equation}
\label{betap eq}
\betap = \frac{\Fdata}{\Nnum}
\end{equation}
Note that 
\[ \beta \le \betap \le  
\beta+\frac{\vsz}{\Nnum \cdot \nsize} + \frac{1}{\Nnum}. \] 
Generally, $\betap \approx \beta$, e.g., for the practical system 
described in Section~\ref{practical setting sec}, 
$\betap \le \beta + \expm{8} + \expm{5}$.

Throughout the remainder of this section, Section~\ref{erepairer sec}
and Section~\ref{core lower sec}, we set
\begin{equation}
\label{mdata eq}
\Mdata = 2 \cdot \Fdata \le \Nnum,
\end{equation}
to be the number of \nfail s in a phase, and thus from Equation~\eqref{betap eq},
\begin{equation}
\label{betap ineq}
\betap \le \frac{1}{2}.
\end{equation}

Note that the restriction  $\betap \le 1/2$ is mild, 
since $\betap \rightarrow 0$ is more interesting 
in practice than $\betap \approx 1$. 

\vspace{0.1in}
\begin{compcorollary}
\label{compression corollary}
Fix any repairer or \lrepairer\ $\Rfnc$, recoverer $\Afnc$,
\timeseq\ $\tseq$, and \didentifierseq\  $\idseq$. Let $\ell \le \dsize$, and
\begin{equation*}
\Xset =
\left\{ \xvar : \rfsize \le \Fdata\cdot \nsize-\ell \right\}
\end{equation*}
where $\rfsize$ is defined with respect to $\xvar$, $\Rfnc$, and $(\tseq,\idseq)$.
\begin{gather}
\label{compression corollary eq}
\Probi{\Xvar}{[\Xvar \in \Xset] \wedge 
[\Afnc(\Sdata(\tvarp_{\Mdata-1})) = \Xvar]} \le 2^{-\ell+1}. \nonumber
\end{gather}
\end{compcorollary}
\begin{proof}
Follows from \Complemma\ and Equations~\eqref{sddata eq}, 
\eqref{sfdata eq}, \eqref{fdata eq}, \eqref{mdata eq}.
\qed
\end{proof}
Note that \Complemma\ and \Compcorollary\ rely upon the
assumption that the \srcdata\ is uniformly distributed, 
and all subsequent technical results rely on this assumption.

A natural relaxation of this assumption is that the
\srcdata\ has high min-entropy, where the min-entropy
is the log base two of one over the probability of $\xvar$, 
where $\xvar$ is the most likely value for the \srcdata.
Thus the min-entropy of the \srcdata\ is always at most $\dsize$.

Since the \srcdata\ for practical systems is composed
of many independent source objects, typically the min-entropy
of the \srcdata\ for a practical system is close to $\dsize$.
It can be verified that all of the lower bounds hold if 
the min-entropy of the \srcdata\ is universally substituted for $\dsize$.

\section{\Erepairer\ lower bound}
\label{erepairer sec}

This section introduces and proves a lower bound
on a constrained repairer, which we hereafter call an \Erepairer, 
within the model introduced in Section~\ref{model sec}.  An
\Erepairer\ is in some ways similar to the \Regrepairer\ of \cite{Dimakis07}, 
\cite{Dimakis10} described in Section~\ref{related work sec}, 
in the sense that between each consecutive \nfail s an \Erepairer\
is constrained to read an equal amount $\frac{\gamma}{\Nnum}$
of data from each of the $\Nnum$ nodes between 
consecutive \nfail s, and thus $\gamma$ is the total amount
of data read from all $\Nnum$ nodes between \nfail s.
Unlike a \Regrepairer, an \Erepairer\ is not constrained in any 
other way, e.g., which data and the amount of data transferred to each
node between \nfail s is unconstrained, and there
is no constraint on when data is transferred to nodes.

\vspace{0.1in}
\begin{elemma}
\label{equal lemma}
For any \Erepairer\ $\Rfnc$ any recoverer $\Afnc$, 
for any \timeseq\ $\tseq$ and \didentifierseq\  $\idseq$,
if
\begin{equation}
\label{eq gam eq}
\gamma \le \frac{\nsize}{2 \cdot \betap}
\end{equation}
then
\begin{gather*}
\nonumber
 \Probi{\Xvar}{\Afnc(\Sdata(\tvarp_{\Mdata-1})) = \Xvar} 
 \le 2^{-\frac{\nsize}{2}+1}.
\end{gather*}

\end{elemma}
\begin{proof}
For any \Erepairer\ $\Rfnc$, 
the amount of data read from each node 
between \nfail s is exactly $\frac{\gamma}{\Nnum}$,
independent of $(\tseq,\idseq)$ and $\xvar$.  Thus, for any $\xvar$,
\begin{gather}
\rfsize = \sum_{i=1}^{\Mdata-1} \frac{i \cdot \gamma}{\Nnum} \\
\le \frac{(\Mdata-1) \cdot \Mdata}{2} \cdot \frac{\nsize}{2 \cdot \betap\cdot \Nnum} \\
\label{equal lb eq}
\le \Fdata\cdot \nsize -\frac{\nsize}{2},
\end{gather}
where Inequality~\eqref{equal lb eq} follows from 
Equations~\eqref{betap eq} and~\eqref{mdata eq}.
The proof follows from applying \Compcorollary\ where 
\[\Xset = \left\{ \xvar  : 
\rfsize \le \Fdata\cdot \nsize-\frac{\nsize}{2}\right\} =\{0,1\}^{\dsize}. \]
\qed
\end{proof}

\section{Core lower bounds}
\label{core lower sec}

From \Compcorollary,  a necessary condition
for \srcdata\ $\xvar$ to be reliably recoverable at the end of the phase
is that repairer or \lrepairer\ $\Rfnc$ must read a lot of data from nodes that fail during the phase, 
and $\Rfnc$ must read this data before the nodes fail.  

On the other hand, $\Rfnc$ cannot predict which nodes are going to fail
during a phase, and only a small fraction of the nodes fail during a phase.
Thus,  to ensure that enough data has been read from nodes 
that fail before the end of the phase, a larger 
amount of data must be read in aggregate from all the nodes.

\Corelemma, the primary technical contribution of this section,
is used to prove \Coretheorem\ and all later results.

\vspace{0.1in}
\begin{corelemma}
\label{core lemma}
Fix $\epscore > 0$ and let 
\begin{equation}
\label{delcore eq}
\delcore =  \Mdata \cdot e^{-\frac{\epscore^2 \cdot \Fdata}{4}+\epscore}.
\end{equation}
For $i = 1, \dots, \Mdata-1$, let 
\begin{equation}
\label{gamma eq}
\Gamma_i = (1-\epscore) \cdot \frac{i \cdot \left(\Nnum-\frac{i+1}{2}\right) \cdot \nsize}
{\Mdata -1}.
\end{equation}
For any repairer or \lrepairer, $\xvar$,  $\tseq$, $\idvar_0$,
\begin{gather*}
\nonumber
 \Probi{\IDseq}{[\forall_{i=1}^{\Mdata -1}\rsize_{i}
 < \Gamma_i] \wedge [\rfsize > (\Fdata-1) \cdot \nsize]} \\ \le \delcore, 
\end{gather*}
where $\IDseq = \ang{\idvar_0, \IDvar_1\ldots,\IDvar_{\Mdata -1}}$.

\end{corelemma}
\begin{proof}
The proof can be found in Appendix~\ref{corelemma proof app}.
\qed
\end{proof}
With the settings in Section~\ref{practical setting sec} and $\betap = 0.1$,
$\delcore \le 3 \cdot 10^{-7}$ when $\epscore = 0.1$,  and
$\delcore \le  10^{-39}$ when $\epscore = 0.2$.

\vspace{0.1in}
\begin{coretheorem}
\label{core theorem}
Fix $\epscore$ with $0 \le \epscore \le 1$,
and Equation~\eqref{delcore eq} defines $\delcore$.
For any repairer $\Rfnc$ and recoverer $\Afnc$, for any fixed $\tseq$, $\idvar_0$,
with probability at most $\delcore + 2^{-\nsize+1}$
with respect to $\Xvar$ and 
$\IDseq = \ang{\idvar_0, \IDvar_1\ldots,\IDvar_{\Mdata -1}}$
the following two statements are both true:
\begin{description}
\item{(1)}
For all $i \in \{1,\ldots,\Mdata-1\}$ the average number of 
\bitval s read by the repairer
between $\tvarm_0$ and $\tvarm_i$ per each of the $i$
\nfail s is less than 
\begin{equation}
 (1-\epscore) \cdot \frac{(1-\betap) \cdot \nsize}{2 \cdot \betap}. \label{core theorem eq}
\end{equation}
\item{(2)}
\Srcdata\ $\Xvar$ is recoverable  at time $\tvarp_{\Mdata-1}$, i.e., 
$\Afnc(\Sdata(\tvarp_{\Mdata-1})) = \Xvar$.
\end{description}
\end{coretheorem}
\begin{proof}
Let
\begin{equation*}
\Xset = \left\{ \xvar : \rfsize \le (\Fdata-1)\cdot \nsize \right\},
\end{equation*}
where $\rfsize$ is defined with respect to $\xvar$, $\Rfnc$, and $\tseq$ and
any $\idseq$, and $\Xbset = \{0,1\}^{\dsize} - \Xset$, i.e.,
\begin{equation*}
\Xbset = \left\{ \xvar : \rfsize > (\Fdata-1)\cdot \nsize \right\}.
\end{equation*}
The probability of (1) and (2) both being true
with respect to $\Xvar$ and $\IDseq$ 
is at most the sum of the following two probabilities 
with respect to $\Xvar$ and $\IDseq$:
\begin{description}
\item({a}) The probability that (1) and (2) and $\Xvar \in \Xset$ are all true. 
This is at most the probability that (2) and $\Xvar \in \Xset$ are both true,
which \Compcorollary\ shows is at most $2^{-\nsize+1}$.
\item({b}) The probability that (1) and (2) and $\Xvar \in \Xbset$ are all true.
This is at most the probability that (1) and $\Xvar \in \Xbset$ are both true.
Note that 
\begin{equation}
\label{gamma ineq}
\frac{\Gamma_i}{i} \ge \frac{\Gamma_{\Mdata-1}}{\Mdata-1} 
\ge (1-\epscore) \cdot \frac{(1-\betap) \cdot \nsize}{2 \cdot \betap}
\end{equation}
for any $i \in \{1,\ldots,\Mdata-1\}$, where Equation~\eqref{gamma eq} 
defines $\Gamma_i$. 
Thus, \Corelemma\ shows that this probability is at most $\delcore$. 
\end{description}
\qed
\end{proof}

Note that $2^{-\nsize}$ is essentially zero in any practical setting.  For example,
$2^{-\nsize}\le 10^{-3 \cdot 10^{15}}$ for the settings in Section~\ref{practical setting sec}.

\section{Main lower bounds}
\label{main sec}

In all previous sections, all the \nfail s within a phase are distinct, 
and when a phase ends and a new phase begins depends 
on the actions of the repairer, and thus the analysis
does not apply to \nfseq s where the \nfail s are
independent of the repairer.  
This section extends the results to random and independent \nfseq s.

\Unitheorem\ in Section~\ref{uniform lower bound sec}
proves lower bounds for any fixed \timeseq\
with respect to a \uidentifierseq.
A \uidentifierseq\ with $\Mdata$ distinct
\nfail s can be generated as follows, where
$\Fdata$ is defined in Equation~\eqref{fdata eq}, and~$\Mdata$
is defined in Equation~\eqref{mdata eq}.

\subsection{\Uidentifierseq\ within a phase}
\label{distinctseq sec}

Let $\Bseq = \{\Bvar(1),\Bvar(2),\ldots \}$ be a
sequence of independently and uniformly distributed in $[0,1]$ random variables.
For $i \ge 1$, define geometric random variable $\Gvar_i$ with respect to $\Bseq$ as
\begin{equation}
\label{geo rv eq}
\Gvar_i = \argmin_{j \ge 1} \left\{ \Bvar(j) \le \frac{\Nnum - i}{\Nnum} \right\},
\end{equation}
and thus $\Exp{\Gvar_i} = \frac{\Nnum}{\Nnum-i}$. 
Let 
\[ \Gseq = \{ \Gvar_1,\ldots,\Gvar_{\Mdata-1} \} \]
be a sequence of independent geometric random variables, each defined
with respect to an independent $\Bseq$ sequence. 
Let 
\[ \IDseq = \ang{\idvar_0,\IDvar_1,\ldots,\IDvar_{\Mdata-1}} \]
be a random \didentifierseq\ as described in Section~\ref{universal sec}.
The \uidentifierseq\ $\Useq$ for the phase 
can be generated as follows from $\Gseq$ and $\IDseq$.
Let $\Gsum_0 = 0$.
For $i = 1,\ldots, \Mdata-1$, let 
\[ \Gsum_i = \sum_{j=1}^i \Gvar_j. \]
For $i \in \{ 1,\ldots,\Mdata-1 \}$, let
\[ \Uvar_{\Gsum_i} = \IDvar_i, \]
and for $j = \Gsum_{i-1}+1,\ldots,\Gsum_i-1$, let
\[ \Uvar_j \inu \{\idvar_0,\IDvar_1,\ldots,\IDvar_{i-1} \}. \]
Then,
\[ \Useq = \{ \idvar_0,\Uvar_1,\Uvar_2,\ldots,\Uvar_{\Gsum_{\Mdata-1}} \} \] 
is a \uidentifierseq.

Note that $\Gsum_0, \Gsum_1,\ldots,\Gsum_{\Mdata-1}$, i.e., the $\Mdata$
indices of the \identifier s in $\Useq$ that are distinct from all the previous \identifier s,
are random variables defined in terms of $\Gseq$.
Thus, $\Gseq$ determines the distinct \nfail\ indices in a phase.
 
For $i = 1,\ldots, \Mdata -1$, let $\Thvar_i = \tvar_{\Gsum_i}$
be the time of the $i^{\rm th}$ distinct \nfail\ beyond the initial \nfail.
 This defines a \timeseq\
 \[ \Tpseq = \{\tvar_0,\Thvar_1,\ldots,\Thvar_{\Mdata -1}\}, \]
which is determined by $(\tseq,\Gseq)$ and is independent of $\IDseq$.

The expected number of \nfail s in a phase until
there are $i$ distinct \nfail s beyond the initial \nfail\ is 
\begin{equation}
\label{exp geosum eq}
\Exp{\Gsum_{i}}
= \sum_{j=1}^{i} \frac{\Nnum}{\Nnum-j}.
\end{equation}
For $0 \le \zeta < 1$, define
\begin{equation}
\label{lnifunction eq}
\lnifunction(\zeta) = \ln\left(\frac{1}{1-\zeta}\right).
\end{equation}
For $0 \le \zeta < 1$,
\begin{equation}
\label{lnifunction ineq}
\sum_{j=0}^{\zeta \cdot \Nnum-1} \frac{1}{\Nnum-j} < \lnifunction(\zeta \cdot \Nnum) < \sum_{j=1}^{\zeta \cdot \Nnum} \frac{1}{\Nnum-j}.
\end{equation}
Setting $\zeta = 2 \cdot \betap$, and using Equations~\eqref{fdata eq},
\eqref{mdata eq}, \eqref{exp geosum eq}, \eqref{lnifunction ineq},
\[ \Exp{\Gsum_{\Mdata-1}} \le \lnifunction(2 \cdot \betap) \cdot \Nnum. \]
Note that as $\zeta \rightarrow 0$,
\begin{equation}
\label{lnifunction asymp eq}
\lnifunction(\zeta) \rightarrow \zeta.
\end{equation}
Thus, $\Exp{\Gsum_{\Mdata-1}} \rightarrow 2 \cdot \betap \cdot \Nnum$
as $\betap \rightarrow 0$.

A phase proceeds as follows with respect to \srcdata\ $\xvar$
and \nfseq\ $(\Tpseq,\IDseq)$,
where Equation~\eqref{gamma eq} defines $\Gamma_i$. 
For $i = 1,\ldots, \Mdata-1$, 
$\Rfnc$ is executed up till time $\Thvarm_i$.  If 
$\rsize_i \ge \Gamma_i$ with respect to $\xvar$, 
$\{\tvar_0,\Thvar_1,\ldots,\Thvar_{i}\}$, 
$\{\idvar_0,\IDvar_1,\ldots,\IDvar_{i-1}\}$
then the phase ends at time $\Thvar_i$.
If the phase doesn't end in the above process then
$\rsize_i < \Gamma_i$ for  $i =1,\ldots,  \Mdata-1$,
and the phase ends at time $\Thvar_{\Mdata-1}$. 

\subsection{Distinct failures lemma} 

The condition $\rsize_i \ge \Gamma_i$ ensures that
that the amount of data read by $\Rfnc$ up till time 
$\Thvar_{\Gsum_{i}}$ in a phase is at least 
$\frac{\Gamma_i}{i}$. However, a lower bound on 
$\frac{\Gamma_i}{\Gsum_{i}}$ is needed, since $\Gsum_{i}$ 
is the total number of \nfail s.  The issue is that $\Gsum_i$ 
is a random variable that can be highly variable relative to $i$
and can depend on $\Rfnc$, and thus $\frac{\Gamma_i}{\Gsum_{i}}$ 
can be highly variable and can be influenced by $\Rfnc$.
Thus, it is difficult to provide lower bounds when considering 
only a single phase.
  
To circumvent these issues,
we stitch phases together into a sequence of phases, and argue that 
$\Rfnc$ must read a lot of data per \nfail\ over a sequence of phases that
covers a large enough number of distinct \nfail s.
\Geolemma\ below proves that if we stitch together enough phases
then we can ensure that, independent of the actions of $\Rfnc$,
with high probability the total number of \nfail s aggregated over 
the phases is close to the expected number of \nfail s relative 
to the number of distinct \nfail s.

The phases can be stitched together as follows.
Let 
 \[ \tseq = \{\tvar_0,\tvar_1,\ldots,\tvar_{i},\ldots\}\]
 be a \timeseq\ and let
\[ \Useq = \{ \idvar_0,\Uvar_1,\Uvar_2,\ldots,\Uvar_{i},\ldots \} \]
be a \uidentifierseq. 
Run $\Rfnc$ on $(\tseq,\Useq)$ until there are $\Mdata$ 
distinct \nfail s in total within the phases, and then continuing running 
$\Rfnc$ until the phase that is underway when there are $\Mdata$ distinct
\nfail s in total completes.  Let $\Yvar$ be the total number of distinct
\nfail s in the phases in this process, where $\Mdata \le \Yvar < 2 \cdot \Mdata$,
and let $\Ypvar$ be the total number of \nfail s in this process, 
where $\Ypvar \ge \Yvar$.  Both $\Yvar$ and $\Ypvar$
are random variables that are determined by $\Rfnc$ and 
$(\tseq,\Useq)$.   

Let 
\begin{equation}
\label{fpdata eq}
\Mpdata = \Mdata \cdot \frac{\lnifunction(2 \cdot \betap)}{2 \cdot \betap} 
= \lnifunction(2 \cdot \betap) \cdot \Nnum.
\end{equation}
From Equation~\eqref{lnifunction asymp eq},
$\Mpdata \rightarrow \Mdata$ as $\betap \rightarrow 0$.

For $0 \le \zeta < 1$, define
\begin{equation}
\label{lndfunction eq}
\lndfunction(\zeta) = \zeta - \ln(1+\zeta).
\end{equation}
Note that as $\zeta \rightarrow 0$,
\begin{equation}
\label{lndfunction asymp eq}
\lndfunction(\zeta) \rightarrow \frac{\zeta^2}{2}.
\end{equation}

\vspace{0.1in}
\begin{geolemma}
\label{geo lemma}
Fix $\betap < 1/2$. Fix $\epsgeo > 0$ and let 
\begin{equation*}
\delgeo =  \Mdata \cdot  
\frac{e^{-2 \cdot \betap \cdot (1-2\cdot \beta') \cdot \Nnum
\cdot \lndfunction(\epsgeo)}}{1+\epsgeo}.
\end{equation*} 
For any repairer $\Rfnc$ and 
recoverer $\Afnc$, for any $\xvar$ and $\tseq$, 
with probability at least $1-\delgeo$ with respect to $\Useq$,
\begin{equation}
\label{geo theorem eq}
\Yvar \ge
\frac{2 \cdot \betap}{(1+\epsgeo) \cdot \lnifunction(2 \cdot \betap)} 
\cdot \Ypvar, 
\end{equation}
and 
\[ \Ypvar \le (1+\epsgeo) \cdot 2 \cdot  \Mpdata. \]

\end{geolemma}

\begin{proof}
The proof can be found in Appendix~\ref{geolemma proof app}.
\qed
\end{proof}

\subsection{Uniform failures lower bound}
\label{uniform lower bound sec}

\vspace{0.1in}
\begin{unitheorem}
\label{uniform theorem}
Fix $\betap < 1/2$. Let $\epscore$ and $\delcore$ be as defined
in \Corelemma, and let $\epsgeo$ and $\delgeo$ be as defined in
\Geolemma, and let 
\[ \deluni = \delgeo + 2 \cdot \Mdata \cdot (\delcore + 2^{-\nsize}). \]
For any repairer $\Rfnc$ and 
recoverer $\Afnc$, for any fixed $\tseq$,
at least one of the following two statements is true
with probability at least  $1-\deluni$
with respect to $\Xvar$ and $\Useq$:
\begin{description}
\item{(1)}
There is an $m \le (1+\epsgeo) \cdot 2 \cdot \Mpdata$ such that the average number of \bitval s read by the repairer between $\tvarm_0$ and $\tvarm_m$ per each of the $m$
\nfail s is at least
\begin{equation}
 \frac{(1-\epscore)}{(1+\epsgeo)} \cdot
 \frac{(1-\betap) \cdot \nsize}{\lnifunction(2 \cdot \betap)}. \label{uniform theorem eq}
\end{equation}
\item{(2)}
\Srcdata\ $\Xvar$ is unrecoverable by $\Afnc$ at time
$\tvarp_{m}$.
\end{description}
\end{unitheorem}
\begin{proof}
From \Geolemma, there is a sequence of phases that ends with 
$\Ypvar \le (1 + \epsgeo) \cdot  2 \cdot \Mpdata$ \nfail s where 
the number of distinct \nfail s $\Yvar$ is at least  
Equation~\eqref{geo theorem eq} with probability at least $1-\delgeo$.
From \Coretheorem\ with respect to all $\xvar$
and $(\tseq,\Gseq,\IDseq)$ and using a union bound 
over at most $2 \cdot \Mdata$ phases in the sequence of phases, 
the average number of \bitval s read
by $\Rfnc$ between $\tvarm_0$ and $\tvarm_{\Ypvar}$ 
per each distinct \nfail\ is at least 
Equation~\eqref{core theorem eq} with probability
at least $1-2 \cdot \Mdata \cdot (\delcore + 2^{-\nsize})$. 
Thus, overall the two statements hold with probability at
least $1-\deluni$.
\qed
\end{proof}

Since the end of one sequence of phases can be the beginning of the next sequence
of phases, it follows that the average number of \bitval s read by the repairer 
per \nfail\ must satisfy Equation~\eqref{uniform theorem eq} 
over the entire lifetime of the system for which the \srcdata\ is recoverable.

Equation~\eqref{uniform theorem eq} holds independent of the \timeseq.
Thus, if there are a lot of \nfail s over a period of time then the
\rrepairrate\ over this period of time must necessarily be high, whereas
if there are fewer \nfail s over a period of time then the \rrepairrate\ over
this period of time can be lower.   Automatic adjustments of the \rrepairrate\
as the \nfail\ rate fluctuates is one of the key contributions of the algorithms
described in \cite{Luby19}, which shows that there are algorithms that can
match the lower bounds of \Unitheorem, even for a fluctuating \timeseq.

\subsection{Poisson failures lower bound}
\label{poisson lower bound sec}

\Unitheorem\ expresses lower bounds in terms of the average number of \bitval s
read per \nfail.  \Poissonthm, presented in this section, instead expresses the lower bounds in terms of a \rrepairrate.  
The primary additional
technical component needed to prove \Poissonthm\ is a concentration
in probability result: the number of \nfail s for a \Ptdist\ with rate $\lambda$
over a suitably long period of time is relatively close to the 
expected number of \nfail s with high probability.

The \Ptdist\ with rate $\lambda$ can be generated as follows. 
For $i \ge 1$, let $\Qvar_i$ be an independent exponential random variable 
with rate $\lambda \cdot \Nnum$,
and let 
\[ \Qseq = \{ \Qvar_1,\ldots,\Qvar_{i}, \ldots \}. \]
For $i \ge 1$, let
\[ \Tvar_i = \tvar_0 + \sum_{j=1}^i \Qvar_j, \]
and let 
\[ \Tseq = \{\tvar_0, \Tvar_1,\ldots,\Tvar_{i}, \ldots \}. \]
For $i \ge 1$, let $\Uvar_i$ be an independent random variable that is
uniformly distributed in $\{0,\ldots,\Nnum-1\},$
and let 
\[ \Useq = \{\idvar_0, \Uvar_1,\ldots,\Uvar_{i}, \ldots \}. \]
Then, $(\Tseq,\Useq)$ is a random \nfseq\ 
with respect to the \Ptdist\ with rate $\lambda$.

Capacity is erased from the system at a rate 
\[ \erate = \lambda \cdot \Nnum \cdot \nsize \]
with respect to the \Ptdist\ with rate $\lambda$. 

\vspace{0.1in}
\begin{Poissontheorem}
\label{Poisson theorem}
Fix $\betap < 1/2$.  Let $\epscore$ be as defined
in \Corelemma, $\epsgeo$ be as defined in \Geolemma,
and $\deluni$ be as defined in \Unitheorem. Let $\epspoi > 0$, let 
\begin{equation*}
\delpoi =  \deluni + (1+\epsgeo) \cdot 2 \cdot \Mpdata 
\cdot  \frac{e^{- \Mdata \cdot \lndfunction(\epspoi) }}{1+\epspoi},
\end{equation*}
and
\begin{equation*}
\Delta = (1+\epsgeo) \cdot (1+\epspoi) \cdot \frac{2 \cdot \lnifunction(2 \cdot \betap)}{\lambda}.
\end{equation*}
For any repairer $\Rfnc$ and 
recoverer $\Afnc$, for any starting time $\tvar_0$,
at least one of the following two statements is true
with probability at least  $1-\delpoi$
with respect to a a \Ptdist\ with rate $\lambda$:
\begin{description}
\item{(1)}
There is a $\tvar \le \tvar_0 + \Delta$ 
such that the average rate $\rrate$ the repairer reads \bitval s between 
$\tvar_0$ and $\tvar$ satisfies
\begin{equation}
\rrate \ge \frac{(1-\epscore)}{(1+\epsgeo) \cdot (1+\epspoi)} \cdot
 \frac{(1-\betap)}{\lnifunction(2 \cdot \betap)} 
 \cdot \erate. \label{Poisson theorem eq}
\end{equation}
\item{(2)}
\Srcdata\ $\Xvar$ is unrecoverable by $\Afnc$ at time $\tvar_0 + \Delta$.
\end{description}
\end{Poissontheorem}
\begin{proof}
From \Unitheorem, there is a sequence of phases that ends with 
$m \le (1 + \epsgeo) \cdot 2 \cdot \Mpdata$ \nfail s where 
the number of distinct \nfail s is provided by 
Equation~\eqref{uniform theorem eq} with probability at least $1-\deluni$.
Since there are at least $\Mdata$ distinct \nfail s in the process, 
$m \ge \Mdata$.

For each $\ell$ between $\Mdata$ and  $(1+\epsgeo) \cdot 2 \cdot \Mpdata$,
when
\[\delpoi' (\ell) = \frac{e^{-\ell \cdot \lndfunction(\epspoi)}}{1+\epspoi}, \]
it follows from Theorem~5.1 of \cite{Janson17} that
\begin{equation*}
\Prob{\sum_{i=1}^{\ell} \Qvar_i \ge (1+\epspoi) \cdot \frac{\ell}{\lambda \cdot N}}
\le \delpoi'(\ell).
\end{equation*}
Using a union bound, it follows that with probability at least 
$1-(1+\epsgeo) \cdot 2 \cdot \Mpdata\cdot \delpoi'(\Mdata)$,
\[
\sum_{i=1}^m \Qvar_i < (1+\epspoi) \cdot \frac{m}{\lambda \cdot N}.
\]
Thus, the time $\tvar = \tvar_0 + \thvar$ when there 
are $m$ \nfail s in the process satisfies
\begin{equation}
\label{time eq}
\thvar \le \frac{(1+\epspoi) \cdot m}{\lambda \cdot \Nnum}
\end{equation}
with probability at least 
$1-(1+\epsgeo) \cdot 2 \cdot \Mpdata\cdot \delpoi'(\Mdata)$.

From \Unitheorem, and combining Equations~\eqref{uniform theorem eq}
and~\eqref{time eq}, it follows that with probability at least $1-\delpoi$
the rate at which the repairer reads data between $\tvar_0$ and $\tvar$
is at least as large as the right-hand side of 
Inequality~\eqref{Poisson theorem eq} or else the \srcdata\
is unrecoverable at time $\tvar = \tvar_0 + \thvar$, 
and thus unrecoverable at time
$\tvar_0 + \Delta \ge  \tvar_0 + \thvar$.
\qed
\end{proof}

From Equation~\eqref{lndfunction asymp eq},
$\delcore$ shrinks exponentially fast as $\Nnum$ 
goes to infinity for fixed $\epscore > 0$,
$\delgeo$ shrinks exponentially fast as $\Nnum$ 
goes to infinity for fixed $\epsgeo > 0$,
and thus $\delpoi$ shrinks exponentially fast as $\Nnum$ 
goes to infinity for fixed $\epscore > 0$, $\epsgeo > 0$, and $\epspoi > 0$.

Since $\epscore>0$, $\epsgeo>0$, and $\epspoi>0$ can be arbitrarily small
constants as $\Nnum$ goes to infinity, the Inequality~\eqref{Poisson theorem eq}
lower bound on $\rrate$ in \Poissonthm\ approaches
\begin{equation}
\label{read rate eq}
\frac{\rrate}{\erate} \ge \frac{1-\betap}{\lnifunction(2 \cdot \betap)}
\end{equation}
as $\Nnum$ goes to infinity.
Since the end of one interval can be the beginning of the next interval, 
it follows that $\rrate$ must also satisfy Equation~\eqref{read rate eq} 
over the entire lifetime of the system.
From Equation~\eqref{lnifunction asymp eq} and Inequality~\eqref{read rate eq},
as $\Nnum$ goes to infinity and $\beta$ goes to $0$, 
the Inequality~\eqref{Poisson theorem eq}
lower bound on $\rrate$ in \Poissonthm\ approaches 
Inequality~\eqref{rrate lb eq}.

\subsection{Distributed storage \srcdata\ capacity}
\label{information sec}

Our distributed storage model and results are inspired 
by Shannon's communication model~\cite{CShannon48}.
For a system with \rcapacity\ $\Nnum \cdot \nsize$,  
we define the \srcdata\ capacity to be
the amount $\dsize$ of \srcdata\ that can be reliably stored 
for long periods of time by the system. 

Based on Equations~\eqref{beta eq} and~\eqref{erate eq},
Inequality~\eqref{rrate lb eq} can be expressed as
\begin{equation}
\label{capacity eq}
\dsize \le \left( 1 - \frac{\erate}{2 \cdot \rrate} \right) \cdot \Nnum \cdot \nsize
\end{equation}
asymptotically as $\Nnum$ and $\frac{\rrate}{\erate}$ approach infinity.  
Inequality~\eqref{capacity eq} expresses a fundamental lower
bound on the \srcdata\ \rcapacity\ as a function of the system \rcapacity,
the \erasurerate\ and the \rrepairrate\ of the repairer.
The paper~\cite{LubyRich19} shows that storage \srcdata\ capacity
asymptotically approaching the righthand side of
Inequality~\eqref{capacity eq} can be achieved
as $\Nnum$ and $\frac{\rrate}{\erate}$ approach infinity,
and thus
\[
\dsize = \left( 1 - \frac{\erate}{2 \cdot \rrate} \right) \cdot \Nnum \cdot \nsize
\] expresses a fundamental 
\srcdata\ \rcapacity\ limit as a function of the system \rcapacity,
the \erasurerate\ and the \rrepairrate\ of the repairer
as $\Nnum$ and $\frac{\rrate}{\erate}$ approach infinity.

\section{Future work}
\label{future sec}

There are many ways to extend this research, accounting for practical issues
in storage system deployments.

Failures in deployed systems can happen at a variable rate that is not known a priori.
For example, a new batch of nodes introduced into a deployment 
may have failure rates that are dramatically different than previous batches.
The paper~\cite{Luby19} introduces repair algorithms that automatically adjust to
fluctuating failure rates.

Both time and spatial failure correlation is common in deployed systems.
Failures in different parts of the system are not completely independent, 
e.g., racks of nodes fail concurrently, entire data centers go offline, 
power and cooling units fail, node outages occur due to rolling 
system maintenance and software updates, etc.
All of these events introduce complicated correlations between
failures of the different components of the system. 

Intermittent \nfail s are common in deployed 
systems, accounting for a vast majority (e.g., 90\%) of \nfail s.  
In the case of an intermittent node failure, the data stored at the node 
is lost for the duration of the failure, but after some period of time
the data stored on the node is available again once the node recovers 
(the period of time can be variable, e.g., ranging from a few seconds to days).
Intermittent failures can also affect entire data centers, a rack of nodes, etc. 

Repairing fragments temporarily unavailable due to 
transient \nfail s wastes network resources.  
Thus, a timer is typically set to trigger a fixed amount of time
after a node fails (e.g., 15 minutes), 
and the node is declared permanently failed and scheduled for repair if it has not recovered within the trigger time. 
Setting the trigger time can be tricky for a \tradsystem; a short trigger time can lead
to unnecessary repair, whereas a long trigger time can reduce reliability.
The paper~\cite{Luby19} provides simulations that highlight the
impact of setting the trigger time for different systems.

Data can silently be corrupted or lost without any notification to the repairer; 
the only mechanism by which a repairer may become aware of such corruption or 
loss of data is by attempting to read the data, i.e., data scrubbing.  
(The data is typically stored with strong checksums, so that the corruption 
or loss of data becomes evident to the repairer
when an attempt to read the data is made.)
For example, the talk~\cite{Cowling16} reports that 
read traffic due to scrubbing can be greater than all other read data traffic combined.
The paper~\cite{Luby19} provides simulations that highlight the
impact of silent corruption on different systems.

There can be a delay between when a node 
permanently fails and when a replacement node is
added.  For example, in many cases adding nodes is performed by robots,
or by manual intervention, and nodes are added in batches instead of individually.

It is important in many systems to distribute the repair 
evenly throughout the nodes and the network,
instead of having a centralized repairer.  
This is important to avoid CPU and network hotspots.
The algorithms described in~\cite{Luby19} distributed the
repair traffic smoothly among all nodes of the system.
The more advanced algorithms 
described in~\cite{LubyRich19}
can be modified to distribute the repair traffic smoothly among all nodes of the system.  
Based on this, it can be seen that distributed versions of the 
lower bounds and upper bounds asymptotically converge as the
\storeoverhead\ approaches zero.

Network topology is an important consideration in deployments, for example when objects are geo-distributed
to multiple data centers.  In these deployments, the available network bandwidth between different nodes
may vary dramatically, e.g., there may be abundant bandwidth available between nodes within the same data center,
but limited bandwidth available between nodes in different data centers.
The paper~\cite{Gopalan16} addresses these issues, 
and the papers~\cite{Gopalan12}, \cite{Huang12}
introduce some erasure codes that may be used in solutions to these issues.
An example of such a deployment is described in~\cite{Facebook14}.

Enhancing the distributed storage model by incorporating the elements described above
into the model and providing an analysis can be of value in understanding fundamental tradeoffs for practical systems.

\section{Conclusions}
We introduce a mathematical model of distributed 
storage that captures some of the relevant features of
practical systems, and prove tight lower bounds on the tradeoff
between the repairer \rrepairrate\ and the \storeoverhead\
as a function of the \erasurerate.
Our hope is that the model and bounds will 
be helpful in understanding and 
designing practical distributed storage systems.

\section*{Acknowledgment}

I thank Roberto Padovani for consistently championing this research.  
I thank members of the Qualcomm systems team 
(Roberto, Tom Richardson, Lorenz Minder, Pooja Aggarwal)
for being great collaborators and for providing invaluable feedback 
on this work as it evolved.
  
I thank the Simons Institute at UC Berkeley for sponsoring 
the Information Theory program January-April 2015, 
as the participants in this program provided a lot of detailed information about work in 
distributed storage that helped understand the context of our research in general.

I thank colleagues at Dropbox (in particular James Cowling), 
Microsoft Azure (in particular Cheng Huang and Parikshit Gopalan), 
Google (in particular Lorenzo Vicisano) and Facebook for sharing valuable
insights into operational aspects and potential issues with large scale deployed distributed storage systems. 
  
I thank Tom Richardson for taking the time to understand and provide 
improvements to this research, including high level presentation suggestions,
simplifications of proofs, and detailed feedback on technical and conceptual inconsistencies.
The feedback from Tom was crucial.

I thank Tom and the organizers of the Shannon lecture series at UCSD for conspiring 
to invite me to give the Shannon lecture December 1, 2015 -- preparing the presentation for this lecture inspired 
thinking about a mathematical model analogous to Shannon's communication theory model.

I thank Chih-Chun Wang for pointing out the deeper connection
between this work and~\cite{Dimakis07}, \cite{Dimakis10},
and in particular for suggesting and providing the proof for  
\Dimlemma\ in Section~\ref{related work sec}.  I also thank
Chih-Chun Wang for extensive comments on reorganizing the
presentation to make it more accessible -- his feedback was invaluable.


%

\begin{IEEEbiographynophoto}{Michael Luby} earned a BSc in Applied Math from MIT and a PhD in Theoretical Computer Science from UC Berkeley. He founded Digital Fountain Inc. in 1999 and served as CTO until acquired by Qualcomm Inc. in 2009, where he was a VP of Technology through late 2018.  He is currently Research Director of the Core Technologies for Transport, Connectivity and Storage Group at the International Computer Science Institute, developing technology which can be leveraged in a variety of transport, communications, and storage solutions, largely based on RaptorQ codes.

Awards for his research include the IEEE Richard W. Hamming Medal, the ACM Paris Kanellakis Theory and Practice Award, the IEEE Eric E. Sumner Communications Theory Award, the UC Berkeley Distinguished Alumni in Computer Science Award, and numerous prizes for his research papers in distributed computing, information theory, coding theory, transport technologies, and cryptography. He is a member of the National Academy of Engineering and is an IEEE Fellow and an ACM Fellow.
\end{IEEEbiographynophoto}

\begin{appendices}

\section{Repairer details}
\label{repairer sec}

This section provides a full description of a repairer,
filling in the details of the brief description provided in 
Section~\ref{repairer initial sec}.

A {\em repairer} $\Rfnc$ can be viewed as a process
that ensures that the \srcdata\ is recoverable 
when data generated from the \srcdata\ is stored at the unreliable nodes.
A repairer for a system operates as follows.
The \identifier\ $\idvar_{i}$ is provided to repairer $\Rfnc$
at time $\tvar_{i}$, which alerts the repairer that all $\nsize$
\bitval s stored on node $\idvar_{i}$ are lost at that time.
As nodes fail and are replaced, the repairer reads data over interfaces from nodes, 
performs computations on the read data, and writes computed data
over interfaces to nodes.
A primary metric is the number of \bitval s the repairer reads over interfaces
from storage nodes.

At time $\tvar$, let
$\Vdata(\tvar)$ be the \bitval s stored in the global memory of $\Rfnc$, 
where $\vsz = \len{\Vdata(\tvar)}$.

Let $\fseq(\tvar)  = (\tseq,\idseq)$ be the \nfseq\ 
up till time $\tvar$, where 
$\tseq = \{\tvar_{0},\ldots,\tvar_{\ell}\},$
$\idseq = \{\idvar_{0},\ldots,\idvar_{\ell}\},$
and $\ell = \argmax_i \{\tvar_i \le t \}$.
The repairer $\Rfnc$ has access to $\fseq(\tvar)$ at time $\tvar$.

The actions of $\Rfnc$ at time $\tvar$ are 
determined by $(\tvar,\Vdata(\tvar),\fseq(\tvar))$.  
If a node $j$ fails at time $\tvar$ then $\Rfnc$ is notified at time $\tvar$ 
that node $j$ failed and $\fseq(\tvar)$ is updated.  If $\Rfnc$ reads data over 
the interface from node $j$ at time $\tvar$ (when to read from node $j$ 
is determined by $(\tvar,\Vdata(\tvar),\fseq(\tvar))$)
then the amount and location of the data read from $\Cdata_j(\tvar)$ 
is determined by $(\tvar,\Vdata(\tvar),\fseq(\tvar))$.
The data read over the interface from node $j$ in response to a request
initiated at time $\tvar$ is assumed to be instantaneously available, 
i.e. all of the requested data is available at time $\tvar$ 
over the interface from node $j$.

$\Vdata(\tvar)$ can be used by the repairer to store 
the programs the repairer executes, store information 
from the past, temporarily store data read from nodes, 
perform computations on read data, temporarily
store computed data before it is written to nodes, and generally to store any
information the repairer needs immediate access to that is not stored at the nodes.  
The distinction between $\Vdata(\tvar)$ and the nodes 
is that $\Vdata(\tvar)$ is persistent memory
 (not subject to any type of failure in the model) and available globally to 
 $\Rfnc$ (there is no read or write cost for accessing $\Vdata(\tvar)$).
 $\Rfnc$ can also store such information at the nodes, but this information
 is subject to loss due to possible \nfail s.

Repairers are allowed to use an unbounded amount of computation,
since computation time is not a metric of interest in the lower bounds.
The granularity of how much data is read or written 
in one step is unconstrained, e.g. one \bitval\ or Terrabytes
of data can be read over an interface from a node during a read step,
and the lower bounds still hold. The granularity of the timing
of read and write steps is also unconstrained, e.g. there may be
a read step each nanosecond, or every twenty minutes. 

{\em \Lrepairer s}, inspired by \cite{Dimakis07} 
and \cite{Dimakis10}, are more powerful than repairers.
The motivation for the \lrepairer\ model
is that a node often has CPUs, memory and storage, 
and often the impact of traffic between storage and memory
at a node is much less than the impact of traffic over the interface
from the node to the system.  Thus, an arbitrary amount of data
may be accessed locally from storage into local memory at a node, local CPUs
may compute and store in the local memory a much smaller amount of data
from the data accessed into local memory, and it is the much smaller amount of
data computed by the CPUs that is sent over the interface from the node to the system. 
The model does not count the data accessed from storage into local memory,
it only counts the data in the local memory that is read by the 
system over the interface from the node.

Formally, for a \lrepairer, when data is to be read over the interface 
from node $j$ initiated at time $\tvar$ (when to read from node $j$ is determined 
by $(\tvar,\Vdata(\tvar),\fseq(\tvar))$), a copy of the entire global memory of the 
\lrepairer\ is assumed to be instantaneously available
 in the local memory at node $j$ at no cost.  As the local computation
 at node $j$ progresses, the copy may evolve to be different than the global
 memory of the \lrepairer\ at time $\tvar$, 
 but the only information the \lrepairer\ potentially receives 
 about any changes to the copy in the local memory 
 is from the locally computed data read 
 by the \lrepairer\ over the interface from node $j$. 
The locally computed data is generated 
based on $(\tvar,\Vdata(\tvar),\fseq(\tvar),\Cdata_j(\tvar))$, 
and then the locally computed data is read by the 
\lrepairer\ over the interface from node $j$.  
The local computational power at node $j$ and the throughput of the interface
at node $j$ are assumed to be unlimited,
and thus the locally computed data requested at time $\tvar$ by the \lrepairer\ is 
available instantly at time $\tvar$ over the interface from node $j$.

Thus the data read over the interface from node $j$ when the request for
the data is initiated at time $\tvar$ is determined by 
$(\tvar,\Vdata(\tvar),\fseq(\tvar),\Cdata_j(\tvar))$.
In this model only the locally computed data is counted 
as data read over the interface from node $j$;  
the data accessed from storage at node $j$ to produce the locally
computed data (which could be all of $\Cdata_j(\tvar)$) is not counted. 

For example, in the extreme a \lrepairer\ could locally access 
all data stored at a node to produce $1$ KB of locally computed data, and 
then only the $1$ KB of locally computed data is read over the interface from
the node.  In this example, only $1$ KB of data is counted towards  
data read by the \lrepairer.  
Thus there is a significant cost to this generalization 
that is not counted in the amount of 
data read from nodes by the repairer.  
These issues are discussed in more detail in Section~\ref{related work sec}.

A repairer is a special case of a \lrepairer: a repairer 
is simply a \lrepairer\ where the data accessed from storage at
the node is directly sent over the interface from the node to the repairer. 

A repairer may employ a randomized algorithm,
which could be modeled by augmenting the repairer
with random and independently chosen \bitval s.   
However, since the repairer is deterministic for 
a fixed setting of the random \bitval s and the lower bounds 
hold for any deterministic repairer, the same lower bounds
hold for any randomized repairer.
Thus, we describe lower bounds only for deterministic repairers, 
noting that all the lower bound results immediately carry over to randomized repairers.

\section{Applying the lower bounds to real systems}
\label{real system sec}

The description of the model makes some very unrealistic assumptions
about how real systems operate in practice.   
However, it is these assumptions that ensure that 
the lower bounds apply to all real systems.
Consider a real system where nodes fail randomly as in the model,
but also portions of the network intermittently fail,
network bandwidth availability is limited and varying
between different parts of the system, 
memory is not completely reliable, multiple distributed semi-autonomous
processes are interacting with different sets of nodes,
responses are not immediate to data requests over
node interfaces,
processes are not immediately notified when nodes fail, 
notification of node failure is not global, 
nodes are not immediately replaced, computational resources are limited, etc.
We describe an omniscient agent acting 
with respect to the model in the role
of the repairer, where the agent emulates the processes
and behaviors of the real system.  
This shows the lower bounds also apply to the real system.

In the model, nodes that fail are immediately 
replaced and the agent is immediately notified.
In the real system, a failed node may not be replaced immediately.
Thus, to emulate the real system, the agent disallows any response to a request to read or write data 
to a failed node from a process until the time 
when the node would have been replaced in the real system. 

In the real system, notifications of node failures may not be instantaneous, and only some processes may be notified.
Thus, the agent only notifies the appropriate processes  
of node failures when they would have been notified in the real system.  

In the model, the agent receives an immediate and complete
response to a request for data over an interface to a node. 
In the real system, interfaces can have a limited amount of 
bandwidth, and there can be delays in delivering responses 
to requests for data by processes over a node interface
due to computational limits or other constraints.
Thus, the agent delivers data to requesting processes 
with the delays and at the speed of the real system.

In the real system, only a small portion of the global
memory state may be available in the local memory
of a node when local-computation repair is used.
Thus, the agent may only need a small portion of the
global memory at the node to emulate a \lrepairer\
of the real system.  

In the model, the agent acting as a repairer has one global memory.
In the real system, repair may be implemented by
a distributed set of processes $\Rfnc_1,\ldots, \Rfnc_i$
executing concurrent reads and writes over node interfaces,
each with their own private memory 
$\Vdata_1(\tvar), \ldots,\Vdata_i(\tvar)$ at time $\tvar$.  
The agent $\Rfnc$ can emulate $\Rfnc_1,\ldots, \Rfnc_i$ as follows.
The global memory of $\Rfnc$ is
\[\Vdata(\tvar) = \{\Vdata_1(\tvar), \ldots,\Vdata_i(\tvar) \}. \] 
If processes $\Rfnc_i$ and $\Rfnc_j$ send \bitval s 
between their local memories at time $\tvar$ then 
these same \bitval s are copied between $\Vdata_i(\tvar)$ 
and $\Vdata_j(\tvar)$ by $\Rfnc$ at time $\tvar$. 
The movement of data between the local memories 
of the processes that the agent is emulating is at no cost.
Thus, the lower bound on the amount of data read over interfaces from 
nodes by $\Rfnc$ in the model is a lower bound on the amount of data 
read over interfaces from nodes by $\Rfnc_1,\ldots, \Rfnc_i$.  

In the model, the agent has a single interface with each node.   
In the real system, a node can have multiple interfaces.
These multiple interfaces are considered as one logical interface by the agent
when counting the amount of data traveling over interfaces from nodes to the agent,
and the agent delivers data traveling over the multiple interfaces to the 
appropriate requesting processes of the emulated real system.

The count of data traffic for the lower bounds is conservative, 
i.e. the amount of data that travels over interfaces from nodes to the agent is a lower bound on the amount 
of data traveling over the network in the real system.

Thus, a real system, whether it is
perfectly architected and has non-failing infinite
network bandwidth, zero computational delays,
 instant node failure notification, or whether it
is more realistic as described above, can be emulated
by the agent in the model as described above. 
Since the lower bounds apply to the agent with respect to the 
model, the lower bounds also apply to any real system.

\section{Proof of \Corelemma}
\label{corelemma proof app}

\begin{proof}
Fix $\xvar$, $\tseq$ and $\idvar_0$. The parameterization with respect to
$\xvar$ and $\tseq$ are implicit in the remainder of the proof. 
Fix $\eta = (\Fdata-1) \cdot \nsize$.
We first prove that for any repairer or \lrepairer\ $\Rpfnc$ there is a repairer or \lrepairer\ $\Rfnc$ such that
\begin{gather}
\label{repairer reduc eq}
\Prob{\left(\forall_{i=1}^{\Mdata -1}\rsize'_{i} < 
\Gamma_i \right) \wedge (\rfsize' > \eta)} \\
\le \Prob{\rfsize > \eta} \nonumber
\end{gather}
and
\begin{equation}
\label{repairer less eq}
\Prob{\forall_{i=1}^{\Mdata -1}\rsize_{i}< \Gamma_i } = 1 
\end{equation}
with respect to $\ang{\idvar_0, \IDvar_1\ldots,\IDvar_{\Mdata -1}}$,
and where $\rsize'_i$ and $\rfsize'$ are defined with respect to $\Rpfnc$ and
$\rsize_i$ and $\rfsize$ are defined with respect to $\Rfnc$.

Let predicate $\Ppred$ be defined as follows on input 
$\ang{\idvar_0,\ldots,\idvar_{\Mdata-1}}$.
\begin{equation*}
\Ppred \mbox{ \rm is true }  \iff \forall_{i=1}^{\Mdata -1}\rsize'_{i}< \Gamma_i 
\end{equation*} 
with respect to $\ang{\idvar_0,\ldots,\idvar_{\Mdata-1}}$.

$\Rfnc$ acts the same way as $\Rpfnc$
with respect to  $\ang{\idvar_0,\ldots,\idvar_{\Mdata-1}}$ 
for which $\Ppred$ is true, thus
$\rsize_i = \rsize'_i$ for $i = 1,\ldots, \Mdata-1$, and $\rfsize = \rfsize'$,
with respect to  $\ang{\idvar_0,\ldots,\idvar_{\Mdata-1}}$
for which $\Ppred$ is true.

Fix  $\ang{\idvar_0,\ldots,\idvar_{\Mdata-1}}$ for which $\Ppred$ is false, let 
\[ \ell = \argmin_{i=1,\ldots,\Mdata-1} \{ \rsize'_i \ge \Gamma_i \} \]
with respect to  $\ang{\idvar_0,\ldots,\idvar_{\Mdata-1}}$.
$\Rfnc$ acts the same way up till time $\tvar_{\ell-1}$, but 
doesn't read data from nodes after $\tvar_{\ell-1}$,
with respect to  $\ang{\idvar_0,\ldots,\idvar_{\Mdata-1}}$.
Thus, $\rsize_i = \rsize'_i$ for $i = 1,\ldots, \ell-1$,
 $\rsize_i = \rsize'_{\ell-1}$ for $i = \ell,\ldots,\Mdata-1$,
with respect to $\ang{\idvar_0,\ldots,\idvar_{\Mdata-1}}$.
From this, $\forall_{i=1}^{\Mdata -1}\rsize_{i}< \Gamma_i$
with respect to any $\ang{\idvar_0,\ldots,\idvar_{\Mdata-1}}$
for which $\Ppred$ is false.  Thus,
condition~\eqref{repairer less eq} holds for repairer $\Rfnc$.

Since $\rfsize = \rfsize'$ with respect to  
all  $\ang{\idvar_0,\ldots,\idvar_{\Mdata-1}}$ for which $\Ppred$ is true,
it follows that
\begin{gather*}
\Prob{\left(\forall_{i=1}^{\Mdata-1}\rsize'_{i} < 
\Gamma_i \right) \wedge (\rfsize' > \eta)} \\
= \Prob{\Ppred = \mbox{ \rm true } \wedge \rfsize > \eta}  \\
\le \Prob{\rfsize > \eta} 
\end{gather*}
with respect to $\ang{\idvar_0, \IDvar_1\ldots,\IDvar_{\Mdata -1}}$,
thus Inequality~\eqref{repairer reduc eq} holds.

The rest of the proof bounds $\Prob{\rfsize > \eta}$
for repairer or \lrepairer\ $\Rfnc$, which provides the bound on 
Inequality~\eqref{repairer reduc eq}.
It can be verified that
\begin{equation}
\label{lenf exp eq}
\Exp{\rfsize_i} = \frac{\rsize_i - \sum_{\ell=1}^{i-1} \rfsize_{\ell}}{\Nnum-i}
\end{equation}
with respect to $\ang{\idvar_0, \ldots\idvar_{i-1},\IDvar_i}$.
Let \[ \rho = \frac{(1-\epscore) \cdot \nsize}{2\cdot \Fdata -1}, \]
\[ \tau_i = \sum_{\ell=1}^i \ell = \frac{i\cdot(i+1)}{2}. \]
If 
\begin{equation}
\label{lenf big eq}
\sum_{\ell=1}^{i-1} \rfsize_\ell \ge \tau_{i-1} \cdot \rho
\end{equation}
with respect to $\ang{\idvar_0, \ldots,\idvar_{i-1}}$
then 
\begin{equation}
\label{lenf bound eq}
\Exp{\rfsize_i} \le i \cdot \rho
\end{equation}
with respect to $\ang{\idvar_0, \ldots,\idvar_{i-1}, \IDvar_i}$.
This follows from Equation~\eqref{lenf exp eq},
Condition~\eqref{repairer less eq}, Inequality~\eqref{lenf big eq},
and because
\[ \frac{\Gamma_i -  \tau_{i-1} \cdot \rho}{\Nnum-i} = i \cdot \rho. \]
Define $\zvar_0=0$, and for $i = 1,\ldots,\Mdata-1$,
\begin{equation}
\label{zdata defn eq}
\Zvar_i  = \zvar_{i-1} + \rfsize_i - i \cdot \rho  = \sum_{\ell=1}^i \rfsize_{\ell} - \tau_i \cdot \rho
\end{equation}
with respect to $\ang{\idvar_0, \ldots,\idvar_{i-1}, \IDvar_i}$,
and define $\zvar_i$ similarly with respect to $\ang{\idvar_0, \ldots,\idvar_{i-1}, \idvar_i}$.
It can be verified that
\[ \tau_{2 \cdot \Fdata -1} \cdot \rho = \Fdata \cdot \nsize - \epscore \cdot \Fdata \cdot \nsize 
= \eta - (\epscore \cdot \Fdata-1) \cdot \nsize, \]
thus 
\begin{equation}
\label{zdata eq}
\Prob{\Zvar_{2 \cdot \Fdata -1} > (\epscore \cdot \Fdata-1) \cdot \nsize} = \Prob{\rfsize > \eta}
\end{equation}
with respect to $\ang{\idvar_0, \IDvar_1,\ldots, \IDvar_{\Mdata - 1}}$.

It can be verified that 
\begin{equation*}
\abs{\zvar_i - \zvar_{i-1}} \le \nsize
\end{equation*}
with respect to all $\ang{\idvar_0, \ldots,\idvar_{i-1}, \idvar_i}$.
Also, Equation~\eqref{zdata defn eq} and 
Inequalities~\eqref{lenf big eq} and \eqref{lenf bound eq} 
imply that if $\zvar_{i-1} \ge 0$ then
\begin{equation*}
\Exp{\Zvar_i} \le \zvar_{i-1}
\end{equation*}
with respect to $\ang{\idvar_0, \ldots,\idvar_{i-1}, \IDvar_i}$.
Thus, $\zvar_0, \Zvar_1,\ldots,\Zvar_{2\cdot \Fdata-1}$ with respect to
$\ang{\idvar_0, \IDvar_1,\ldots, \IDvar_{\Mdata -1}}$ satisfies the conditions
of \Supertheorem\ of Appendix~\ref{super sec} 
with $n=\Mdata-1$, $c = \nsize$, 
and $\alpha = (\epscore \cdot \Fdata - 1) \cdot \nsize$.
Thus, from \Supertheorem\
and Equation~\eqref{zdata eq}, it can be verified that
\[ \Prob{\rfsize > \eta} \le \delcore. \]
 with respect to
$\ang{\idvar_0, \IDvar_1,\ldots, \IDvar_{\Mdata -1}}$.
The lemma follows from Inequality~\eqref{repairer reduc eq}.
\qed
\end{proof} 

\section{\Supertheorem}
\label{super sec}

We provide a probability bound used in the proof of 
\Corelemma\
that may be of independent interest.  Any improvement to this bound
provides an immediate improvement to \Corelemma.
We generalize previous notation. 
\vspace{0.1in}
\begin{supertheorem}
\label{super theorem}  
Let $\zvar_0, \Zvar_1,\ldots,\Zvar_{n}$ 
be a random sequence of real-values
defined with respect to another random sequence
$\set{\idvar_0, \IDvar_1,\ldots, \IDvar_{n}}$, such that $z_0=0$ and the 
following conditions are satisfied for $i = 1,\ldots, n$.
\begin{itemize}
\item
$z_i$ is determined by $\set{\idvar_0, \idvar_1,\ldots, \idvar_{i}}$.
\item
$\abs{\zvar_i - \zvar_{i-1}} \le c$
with respect to all $\set{\idvar_0, \ldots,\idvar_{i-1}, \idvar_i}$.
\item
if $\zvar_{i-1} >  0$ then $\Exp{\Zvar_i} \le \zvar_{i-1}$
with respect to $\set{\idvar_0, \ldots,\idvar_{i-1}, \IDvar_i}$.
\end{itemize}
Then, for any $\alpha > 0$,
\begin{equation*}
\Prob{\Zvar_{n} > \alpha+c} \le n \cdot e^{\frac{-\alpha^2}{2 \cdot n \cdot c^2}}
\end{equation*}
\end{supertheorem}
\begin{proof}
For $\ell= 1,\ldots,n$, let predicate $\Ppred_\ell$ be
defined as follows on input 
$\set{\idvar_0,\ldots,\idvar_\ell,\ldots,\idvar_i}$, 
with $i \in \{ \ell,\ldots,n \}$. 
\begin{equation*}
\Ppred_\ell \mbox{ \rm is true }  \iff \zvar_{\ell-1} \le 0 \wedge \zvar_\ell > 0
\end{equation*} 
with respect to $\set{\idvar_0,\ldots,\idvar_\ell,\ldots,\idvar_i}$.  

For each $\ell= 1,\ldots,n$ and each $\set{\idvar_0,\ldots,\idvar_\ell}$ such
that $\Ppred_\ell$ is true, define a sequence as follows.
\begin{itemize}
\item
$\zvar^{\ell,\set{\idvar_0,\ldots,\idvar_\ell}}_\ell = \zvar_\ell$ 
with respect to $\set{\idvar_0,\ldots,\idvar_\ell}$.
\item
For $i = \ell+1,\ldots, n$, 
\begin{align}
\Zvar^{\ell,\set{\idvar_0,\ldots,\idvar_\ell}}_i & = \Zvar_{i} & 
{\rm if } \zvar^{\ell,\set{\idvar_0,\ldots,\idvar_\ell}}_{i-1} > 0 \label{super 1 eq}\\
\Zvar^{\ell,\set{\idvar_0,\ldots,\idvar_\ell}}_i & = 
\zvar^{\ell,\set{\idvar_0,\ldots,\idvar_\ell}}_{i-1} & 
{\rm if } \zvar^{\ell,\set{\idvar_0,\ldots,\idvar_\ell}}_{i-1} \le 0 \label{super 2 eq}
\end{align}
with respect to 
$\set{\idvar_0, \ldots,\idvar_\ell, \idvar_{\ell+1},\ldots, \idvar_{i-1}, \IDvar_i}$.
\end{itemize}
It can be verified that, for all 
$\set{\idvar_0, \ldots,\idvar_\ell, \idvar_{\ell+1},\ldots, \idvar_i}$,
\begin{equation}
\label{super bound eq}
\abs{\zvar^{\ell,\set{\idvar_0,\ldots,\idvar_\ell}}_i - 
\zvar^{\ell,\set{\idvar_0,\ldots,\idvar_\ell}}_{i-1}} \le c
\end{equation}
with respect to $\set{\idvar_0, \ldots,\idvar_\ell, \idvar_{\ell+1},\ldots, \idvar_i}$.

With respect to $\set{\idvar_0, \ldots,\idvar_\ell, \idvar_{\ell+1},\ldots, \idvar_{i-1},\IDvar_i}$:
Equations~\eqref{super 1 eq} and~\eqref{super 2 eq} imply 
that $\zvar^{\ell,\set{\idvar_0,\ldots,\idvar_\ell}}_{i-1} =  \zvar_{i-1}$
if $\zvar^{\ell,\set{\idvar_0,\ldots,\idvar_\ell}}_{i-1} > 0$, 
and since $\Exp{\Zvar_i} \le \zvar_{i-1}$ if $\zvar_{i-1} > 0$, it follows that
\[ \Exp{\Zvar^{\ell,\set{\idvar_0,\ldots,\idvar_\ell}}_i} \le 
\zvar^{\ell,\set{\idvar_0,\ldots,\idvar_\ell}}_{i-1} \]
if $\zvar^{\ell,\set{\idvar_0,\ldots,\idvar_\ell}}_{i-1} > 0$.
From Equation~\eqref{super 2 eq}, 
\[ \Exp{\Zvar^{\ell,\set{\idvar_0,\ldots,\idvar_\ell}}_i} = \zvar^{\ell,\set{\idvar_0,\ldots,\idvar_\ell}}_{i-1} \] 
if $\zvar^{\ell,\set{\idvar_0,\ldots,\idvar_\ell}}_{i-1} \le 0$.
Thus, 
\begin{equation}
\label{super exp eq}
\Exp{\Zvar^{\ell,\set{\idvar_0,\ldots,\idvar_\ell}}_i} \le  
\zvar^{\ell,\set{\idvar_0,\ldots,\idvar_\ell}}_{i-1}
\end{equation}
with respect to 
$\set{\idvar_0, \ldots,\idvar_\ell, \idvar_{\ell+1},\ldots, \idvar_{i-1},\IDvar_i}$.

From Equations~\eqref{super bound eq} and~\eqref{super exp eq}, 
for $\ell = 1,\ldots, n$, for all $\set{\idvar_0, \ldots, \idvar_{\ell}}$ 
such that $\Ppred_\ell$ is true,
\[ \zvar^{\ell,\set{\idvar_0,\ldots,\idvar_\ell}}_\ell, 
\Zvar^{\ell,\set{\idvar_0,\ldots,\idvar_\ell}}_{\ell+1},\ldots,
\Zvar^{\ell,\set{\idvar_0,\ldots,\idvar_\ell}}_{n} \] 
with respect to
$\set{\idvar_0, \ldots, \idvar_\ell, \IDvar_{\ell+1}, \ldots, \IDvar_{n}}$
is a supermartingale.
Thus, from the Azuma's inequality, 
\begin{equation}
\label{azuma eq}
\Prob{\Zvar^{\ell,\set{\idvar_0,\ldots,\idvar_\ell}}_{n} - \zvar^{\ell,\set{\idvar_0,\ldots,\idvar_\ell}}_\ell > \alpha} \le e^{\frac{-\alpha^2}{2 \cdot (n-\ell) \cdot c^2}}
\end{equation}
with respect to $\set{\idvar_0, \ldots,\idvar_\ell, \IDvar_{\ell+1},\ldots, \IDvar_{n}}$.
It can be verified that $\zvar^{\ell,\set{\idvar_0,\ldots,\idvar_\ell}}_\ell \le c$ if $\Ppred_\ell$ is true for 
$\set{\idvar_0,\ldots,\idvar_\ell}$, thus
\begin{gather}
\label{martin corr eq}
\Prob{\Zvar^{\ell,\set{\idvar_0,\ldots,\idvar_\ell}}_{n} > \alpha + c} \\
\le \Prob{\Zvar^{\ell,\set{\idvar_0,\ldots,\idvar_\ell}}_{n} - \zvar^{\ell,\set{\idvar_0,\ldots,\idvar_\ell}}_\ell > \alpha} \nonumber
\end{gather}
with respect to $\set{\idvar_0, \ldots,\idvar_\ell, \IDvar_{\ell+1},\ldots, \IDvar_{n}}$.

It can be verified that, for any 
$\set{\idvar_0, \ldots, \idvar_{n}}$,
\begin{gather}
\label{super corr eq}
\zvar_n > \alpha + c \iff \\
\exists_{\ell = 1}^{n} \mbox{ \rm s.t. }
\Ppred_\ell \mbox{ \rm is true } \wedge 
\zvar^{\ell,\set{\idvar_0,\ldots,\idvar_\ell}} > \alpha + c \nonumber
\end{gather}
with respect to $\set{\idvar_0, \ldots, \idvar_{n}}$.
From Equation~\eqref{super corr eq} it follows that
\begin{gather}
\label{giant eq}
\Prob{\Zvar_n > \alpha + c} \le \\ 
\sum_{\ell=1}^n \Prob{\Ppred_\ell \mbox{ \rm is true }  \wedge 
\Zvar^{\ell,\set{\idvar_0, \IDvar_1,\ldots, \IDvar_{\ell}}}_n > \alpha+c} \nonumber \\
\end{gather}
with respect to $\set{\idvar_0,\IDvar_1, \ldots, \IDvar_{n}}$.
From Inequalites~\eqref{giant eq}, \eqref{martin corr eq}, \eqref{azuma eq},
it follows that 
\begin{gather}
\Prob{\Zvar_n > \alpha + c} \le  \sum_{\ell=1}^n 
e^{\frac{-\alpha^2}{2 \cdot (n-\ell) \cdot c^2}} \le 
n \cdot e^{\frac{-\alpha^2}{2 \cdot n \cdot c^2}},
\end{gather}
with respect to $\set{\idvar_0,\IDvar_1, \ldots, \IDvar_{n}}$.
\qed
\end{proof}

\section{Proof of \Geolemma}
\label{geolemma proof app}

\begin{proof}
For now, fix $\ell$ with $\Mdata \le \ell \le 2 \cdot \Mdata$.
Run $\Rfnc$ with respect to $x$, $\tseq$ and $\Useq$ until
the aggregate number of distinct \nfail s in the phases is exactly $\ell$
(which may occur in the middle of an uncompleted phase).
Let $\pvar$ be the number of phases including the last possibly partially completed phase.
For $j=1,\ldots,\pvar$, let $\dvar_j$ be the number of 
distinct \nfail s in phase $j$.  
As described in Section~\ref{distinctseq sec}, let
\[ \Gseq = \{ \Gvar^1_1,\ldots,\Gvar^1_{\dvar_1}, \Gvar^2_1,\ldots,\Gvar^2_{\dvar_2}, \ldots, \Gvar^\pvar_1,\ldots,\Gvar^\pvar_{\dvar_\pvar}\} \]
be the independent geometric random variables used in the $\pvar$ phases,
where $\Gvar^j_i$ is the same as $\Gvar_i$ defined in Equation~\eqref{geo rv eq}.
Note that $\sum_{j=1}^\pvar \dvar_j = \ell$, and
\[ \Yppvar = \sum_{j=1}^{\pvar} \sum_{i=1}^{\dvar_j} \Gvar^j_i \]
is the number of \nfail s in the sequence of phases.

The random variables in $\Gseq$ depend on the
history of the process. For example, $\Gvar^j_i$ is used to determine
the index of the next distinct \nfail\ after there are $i-1$ distinct \nfail s
in a phase $j$ that has not yet terminated. 
The geometric random variable in the sequence after $\Gvar^j_i$
depends on the actions of $\Rfnc$ up till the time of the next distinct
\nfail\ in phase $j$, where the actions of $\Rfnc$ may depend 
on the evolving value of $\Gvar^j_i$ during this time.  
If the actions of $\Rfnc$ cause phase $j$ not to terminate
after there are $i$ distinct \nfail s in phase $j$ then the next geometric
random variable in the sequence is $\Gvar^j_{i+1}$, 
whereas if the actions of $\Rfnc$ cause phase $j$ to terminate
after there are $i$ distinct \nfail s in phase $j$ then the next 
geometric random variable in the sequence is $\Gvar^{j+1}_{1}$.

However, once which geometric random variable to use next
is determined within $\Gseq$, the value of the determined 
geometric random variable is chosen independently of all previous 
history of the process, i.e., independent of the previous geometric
random variables and their values in $\Gseq$, and independently of $\Rfnc$.
Thus, $\Gseq$ is a sequence of independent random variables,
but which random variables are in $\Gseq$ depends on the process.

Let $\phvar = \lceil \frac{\ell}{\Mdata-1}\rceil,$
for $j = 1,\ldots,\phvar-1$ let $\dhvar_j = \Mdata-1$,
let $\dhvar_{\phvar} = \ell- (\Mdata-1) \cdot (\phvar-1)$, 
For  $j = 1,\ldots,\phvar$, $i=1,\ldots,\dhvar_j$, let 
\[ \Bseq^j_i = \{ \Bvar^j_i(1), \Bvar^j_i(2), \ldots \} \] 
be a sequence of independently and uniformly 
distributed in $[0,1]$ random variables, and let
\[ \Bseqseq = \{ \Bseq^j_i :  j = 1,\ldots,\phvar, i=1,\ldots,\dhvar_j \} \]
be a sequence of $\ell$ such sequences. Let 
\[ \Ghseq = \{ \Ghvar^1_1,\ldots,\Ghvar^1_{\dhvar_1}, \Ghvar^2_1,\ldots,\Ghvar^2_{\dhvar_2}, \ldots, \Ghvar^{\phvar}_1,\ldots,\Ghvar^{\phvar}_{\dhvar_{\phvar}},\} \]
where $\Ghvar^j_i$ is calculated using $\Bseq^j_i$ as described in Section~\ref{distinctseq sec}
and defined in Equation~\eqref{geo rv eq}.

As the sequence $\Gseq = \{ \Gseq^1, \Gseq^2, \Gseq^\pvar \}$ of $\ell$ geometric
random variables defined by the process above is being generated,
after index $i$ of phase $j$ has been determined, $\Gvar^j_i$ 
can be matched with an unmatched $\Ghvar^{j'}_{i'}$ of $\Ghseq$,
where 
\[ i' = \argmin_{i'} \set{ i' \ge i: \exists_{j'} \mbox{ s.t. } 
 \Ghvar^{j'}_{i'} \mbox{ is unmatched}}, \] 
 and thus $\Gvar^j_i$ is matched with $\Ghvar^{j'}_{i'}$ where $i' \ge i$.
There is always a match because, for all $i$, 
\[ \abs{\set{\Gvar^j_{i'} \in \Gseq \mbox{ s.t. } i' \ge i}} 
\le \abs{\set{\Ghvar^j_{i'} \in \Ghseq \mbox{ s.t. } i' \ge i}}. \]
This holds independent of which random
variables are added to $\Gseq$ by the process. 
Only a prefix of $\Gseq$ is known at the time of each 
match, but all of $\Ghseq$ is known a priori.

If $\Gvar^j_i$ is matched to $\Ghvar^{j'}_{i'}$ then the value of $\Gvar^j_i$
can be calculated as described in Section~\ref{distinctseq sec}
and defined in Equation~\eqref{geo rv eq} using the same $\Bseq^{j'}_{i'}$
as is used to calculate $\Ghvar^{j'}_{i'}$.
From $i' \ge i$ it follows that $\Ghvar^{j'}_{i'} \ge \Gvar^j_i$ 
for all possible values of the random variables in $\Bseq^{j'}_{i'}$.
Thus, $\Ghseq$ is determined by $\Bseqseq$, $\Gseq$ is determined
by $\Rfnc$, $\tseq$ and $\Bseqseq$, and,  for any $\Rfnc$, $\xvar$ and $\tseq$, for any positive $\eta$,
\begin{gather}
\label{dom eq} 
\Probi{\Bseqseq}{\Yppvar \ge \eta}
\le  \Probi{\Bseqseq}{\sum_{j=1}^{\phvar} \sum_{i=1}^{\dhvar_j} \Ghvar^j_i \ge \eta}.
\end{gather}
It can be verified from Equation~\eqref{lnifunction ineq} that 
\begin{equation}
\label{geosum eq}
\Exp{\sum_{j=1}^{\phvar} \sum_{i=1}^{\dhvar_j} \Ghvar^j_i} \le 
\frac{\lnifunction(2 \cdot \betap)}{2 \cdot \betap}\cdot \ell.
\end{equation}
Let 
\begin{equation*}
\delgeo' = \frac{e^{-2 \cdot \betap \cdot (1-2\cdot \beta') \cdot \Nnum
\cdot \lndfunction(\epsgeo)}}{1+\epsgeo}.
\end{equation*}
From Inequality~\eqref{geosum eq}, since the lefthand sum in Inequality~\eqref{geosum eq}
is over $\ell \ge \Mdata = 2 \cdot \betap \cdot \Nnum$ geometric random variables, 
Theorem 2.1 of~\cite{Janson17} implies that
\begin{equation}
\label{geo eq}
\Probi{\Bseqseq}{\sum_{j=1}^{\phvar} \sum_{i=1}^{\dhvar_j} \Ghvar^j_i  
\ge (1+\epsgeo) \cdot \frac{\lnifunction(2 \cdot \betap)}{2 \cdot \betap}\cdot \ell } \le \delgeo',
\end{equation}
and from Inequalities~\eqref{dom eq} and~\eqref{geo eq} it follows that, for any $\Rfnc$, $\xvar$ and $\tseq$,
\begin{equation}
\label{ellep eq}
\Probi{\Bseqseq}{\Yppvar  \ge (1+\epsgeo) \cdot \frac{\lnifunction(2 \cdot \betap)}{2 \cdot \betap}\cdot \ell} \le \delgeo'.
\end{equation}

Now consider the process described just prior to the statement of
\Geolemma, where random variable $\Yvar$ 
is the number of distinct \nfail s
in the sequence of phases and random variable $\Ypvar$ 
is the number of overall \nfail s.  
Let $\delgeo = \Mdata \cdot \delgeo'$.
Since there are at most $\Mdata$ possible values for $\Yvar$, 
Inequality~\eqref{ellep eq} and a union bound show that,
for any $\Rfnc$, $\xvar$ and $\tseq$, 
\[ \Probi{\Useq}{\Ypvar \ge (1+\epsgeo) \cdot 
\frac{\lnifunction(2 \cdot \betap)}{2 \cdot \betap} \cdot \Yvar} \le \delgeo, \]
and thus
\[ \Probi{\Useq}{\Yvar \ge \frac{2 \cdot \betap} {(1+\epsgeo) 
\cdot \lnifunction(2 \cdot \betap)} \cdot \Ypvar} \ge 1-\delgeo. \]
This also shows that
\[ \Probi{\Useq}{\Ypvar \le  (1+\epsgeo) \cdot 2 \cdot \Mpdata} \ge 1-\delgeo, \]
since 
$\Yvar \le 2 \cdot \Mdata = 2 \cdot \Mpdata \cdot \frac{2 \cdot \betap}{\lnifunction(2 \cdot \betap)}$ from Equation~\eqref{fpdata eq}.  
\qed
\end{proof}

\end{appendices}








\end{document}